\documentclass[amsmath,amssymb,floatfix,lengthcheck,superscriptaddress,longbibliography]{revtex4-1}

\usepackage{graphicx}
\usepackage[centering,hmargin=20mm,tmargin=29mm,bmargin=25mm]{geometry}
\usepackage{multirow}
\usepackage{newtxtext}
\usepackage[cmintegrals]{newtxmath} 
\usepackage{bm}
\usepackage{bbm}
\usepackage{etoolbox}
\usepackage{setspace}
\usepackage{subfigure}
\usepackage{titlesec}
\usepackage{bibunits}

\titleformat{\section}{\raggedright\bfseries}{\arabic{section}.}{1em}{}

\usepackage{xcolor}
\usepackage[pdfencoding=auto]{hyperref}
\usepackage{cleveref}
\hypersetup{colorlinks,
	linkcolor={blue!75!black!80!yellow},
	citecolor={blue!75!black!80!yellow},
	urlcolor={blue!75!black!80!yellow}
}

\makeatletter
\renewcommand\@make@capt@title[2]{%
	\@ifx@empty\float@link{\@firstofone}{\expandafter\href\expandafter{\float@link}}%
	\sffamily{\textbf{#1}}\@caption@fignum@sep#2
}%

\makeatother

\crefname{Fig}{Fig.}{Figs.}
\Crefname{Fig}{Figure.}{Figures.}

\defaultbibliography{wte2_hydro_citations}
\defaultbibliographystyle{apsrev4-1}

\begin{document}

\title{Imaging phonon-mediated hydrodynamic flow in WTe\texorpdfstring{\textsubscript{2}}{\texttwoinferior}} 
\author{Uri Vool}
\thanks{These authors contributed equally.}
\affiliation{John Harvard Distinguished Science Fellows Program, Harvard University, Cambridge, MA 02138, USA}
\affiliation{Department of Physics, Harvard University, Cambridge, MA 02138, USA}
\author{Assaf Hamo}
\thanks{These authors contributed equally.}
\affiliation{Department of Physics, Harvard University, Cambridge, MA 02138, USA}
\author{Georgios Varnavides}
\thanks{These authors contributed equally.}
\affiliation{John A. Paulson School of Engineering and Applied Sciences, Harvard University, Cambridge, MA 02138, USA}
\affiliation{Department of Materials Science and Engineering, Massachusetts Institute of Technology, Cambridge, MA 02139, USA}
\affiliation{Research Laboratory of Electronics, Massachusetts Institute of Technology, Cambridge, MA 02139, USA}
\author{Yaxian Wang}
\thanks{These authors contributed equally.}
\affiliation{John A. Paulson School of Engineering and Applied Sciences, Harvard University, Cambridge, MA 02138, USA}
\author{Tony X. Zhou}
\affiliation{Department of Physics, Harvard University, Cambridge, MA 02138, USA}
\affiliation{John A. Paulson School of Engineering and Applied Sciences, Harvard University, Cambridge, MA 02138, USA}
\author{Nitesh Kumar}
\affiliation{Max-Planck-Institut für Chemische Physik Fester Stoffe, Dresden, Germany}
\author{Yuliya Dovzhenko}
\affiliation{Department of Physics, Harvard University, Cambridge, MA 02138, USA}
\author{Ziwei Qiu}
\affiliation{Department of Physics, Harvard University, Cambridge, MA 02138, USA}
\affiliation{John A. Paulson School of Engineering and Applied Sciences, Harvard University, Cambridge, MA 02138, USA}
\author{Christina A. C. Garcia}
\affiliation{John A. Paulson School of Engineering and Applied Sciences, Harvard University, Cambridge, MA 02138, USA}
\author{Andrew T. Pierce}
\affiliation{Department of Physics, Harvard University, Cambridge, MA 02138, USA}
\author{Johannes Gooth}
\affiliation{Department of Physics, Harvard University, Cambridge, MA 02138, USA}
\affiliation{Max-Planck-Institut für Chemische Physik Fester Stoffe, Dresden, Germany}
\affiliation{Institut für Festkörper- und Materialphysik, Technische Universität Dresden, 01062 Dresden, Germany}
\author{Polina Anikeeva}
\affiliation{Department of Materials Science and Engineering, Massachusetts Institute of Technology, Cambridge, MA 02139, USA}
\affiliation{Research Laboratory of Electronics, Massachusetts Institute of Technology, Cambridge, MA 02139, USA}
\author{Claudia Felser}
\affiliation{John A. Paulson School of Engineering and Applied Sciences, Harvard University, Cambridge, MA 02138, USA}
\affiliation{Max-Planck-Institut für Chemische Physik Fester Stoffe, Dresden, Germany}
\author{Prineha Narang}
\email[Electronic address:\;]{prineha@seas.harvard.edu}
\affiliation{John A. Paulson School of Engineering and Applied Sciences, Harvard University, Cambridge, MA 02138, USA}
\author{Amir Yacoby}
\email[Electronic address:\;]{yacoby@g.harvard.edu}
\affiliation{Department of Physics, Harvard University, Cambridge, MA 02138, USA}
\affiliation{John A. Paulson School of Engineering and Applied Sciences, Harvard University, Cambridge, MA 02138, USA}
\date{\today}

\maketitle 
\begin{bibunit}
\textbf{
In the presence of interactions, electrons in condensed matter systems can behave hydrodynamically, exhibiting phenomena associated with classical fluids, such as vortices and Poiseuille flow~\cite{gurzhi_hydrodynamic_1968,levitov_electron_2016,Lucas_2018}.
In most conductors, the electron-electron interactions are minimised by screening effects, hindering the search for hydrodynamic materials, but recently a class of semimetals has been reported to exhibit prominent interactions~\cite{moll_evidence_2016,gooth_thermal_2018}.
In this work, we study the current flow in the layered semimetal tungsten ditelluride (WTe$_2$) by imaging the local magnetic field using a nitrogen-vacancy defect in diamond. 
We image the spatial current profile within three-dimensional WTe$_2$ and find it exhibits nonuniform current density, indicating hydrodynamic flow. 
Our temperature-resolved current profile measurements reveal a non-monotonic temperature dependence, with hydrodynamic effects strongest at approximately  $20~\mathrm{K}$. 
We also report \textit{ab initio} calculations that show the electron-electron interactions are not explained by the Coulomb interaction alone, but are predominantly mediated by phonons. 
This provides a promising avenue in the search for hydrodynamic flow and prominent electron interactions in high carrier density materials.
}

When microscopic scattering processes in condensed matter systems conserve momentum, electrons flow collectively, akin to a fluid, deviating significantly from the expected diffusive flow of a Fermi liquid.
This behavior, termed hydrodynamic electron flow, has been reported in transport measurements in (Al,Ga)As~\cite{de_jong_hydrodynamic_1995}, graphene~\cite{bandurin_negative_2016,crossno_observation_2016,krishna_kumar_superballistic_2017}, PdCoO$_2$~\cite{moll_evidence_2016}, and WP$_2$~\cite{gooth_thermal_2018}. 
Even when momentum is conserved, the presence of boundaries does not need to respect this conservation, leading to a distinct spatial signature when current flows along a channel, known as Poiseuille flow. 
The resulting current profile is characterized by enhanced flow in the center of the channel, and reduced flow along the edges, as has recently been shown by spatially-resolved measurements in graphene~\cite{sulpizio_visualizing_2019,ku_imaging_2020,jenkins_imaging_2020}.
In bulk hydrodynamically-reported materials (such as PdCoO$_2$~\cite{moll_evidence_2016}, and WP$_2$~\cite{gooth_thermal_2018}), however, significant carrier density presents unsolved mysteries regarding the microscopic origins of hydrodynamics.

Tungsten ditelluride (WTe$_2$) is a layered semimetal which has stirred significant interest in recent years as part of a new class of quantum materials.
The bulk crystal exhibits large magnetoresistance~\cite{ali_large_2014} and pressure-driven superconductivity~\cite{kang_superconductivity_2015,pan_pressure-driven_2015}, and has been predicted~\cite{soluyanov_type-ii_2015} and observed~\cite{li_evidence_2017,lin_visualizing_2017,lv_experimental_2017,sie_ultrafast_2019} to be a Type-II Weyl semimetal. 
In the monolayer, WTe$_2$ can be electrostatically gated into a quantum spin Hall insulator~\cite{wu_observation_2018} or a superconductor~\cite{fatemi_electrically_2018,sajadi_gate-induced_2018}.
These effects are due to the rich band structure of the material, its high conductance (carrier mean-free-paths as long as $12~\mathrm{\mu m}$~\cite{zhu_quantum_2015}), and significant electron-electron interactions.
Since momentum is conserved during electron-electron interactions, spatially resolved measurements of hydrodynamic flow in WTe$_2$ are uniquely positioned to probe interactions in the material.

\begin{figure*}
\centering
\includegraphics[width=\textwidth]{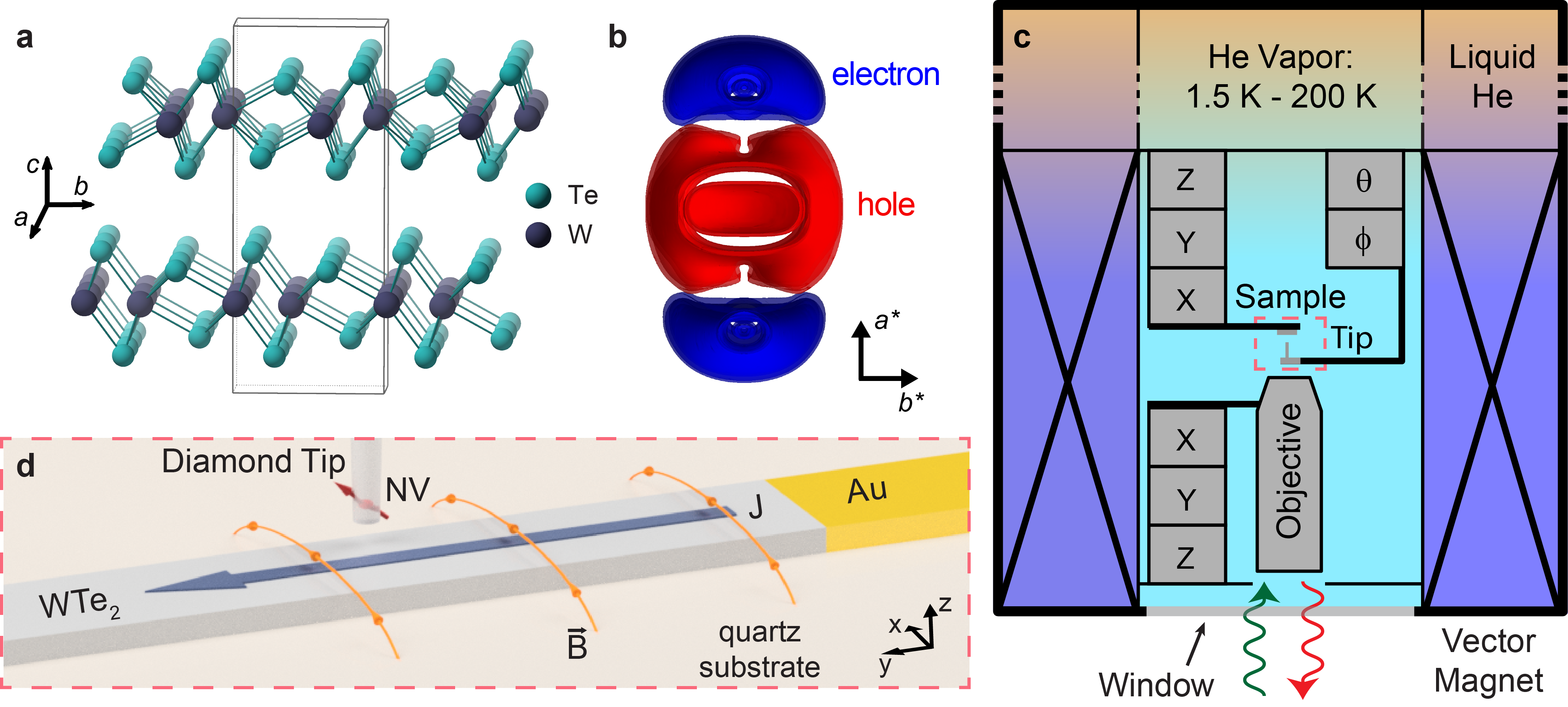}
\caption{\textbf{Device and experimental setup.} 
(\textbf{a}) Crystal structure highlighting the layered-structure of WTe$_2$. 
(\textbf{b}) Calculated Fermi surface for the WTe$_2$ lattice, with compensated electron (blue) and hole (red) pockets.
(\textbf{c}) Scanning magnetometry microscope mounted in a flow cryostat. 
The outer layer is filled with liquid He and houses the vector magnet. 
A needle valve connects this bath to the central chamber, allowing for the flow of He vapor into it to control temperature by balancing flow rate and heating. 
The microscope consists of a diamond tip with a NV defect in contact with the sample, which can be moved by piezo-electric controllers.
The diamond tip is attached to goniometers which allow for angle-control (see Methods). 
The bottom of the cryostat has a window for optical access, allowing us to measure the NV defect's spin state.
The objective, which focuses the light on the defect, can also be moved.
(\textbf{d}) A closeup of the diamond tip with the NV defect, and the WTe$_2$ flake. 
Current is flowing through the flake, generating magnetic field which is measured by the NV. }
\end{figure*}

In this \emph{Letter}, we present the spatial profile of current flowing in a WTe$_2$ flake and show it is consistent with hydrodynamic flow; this is the first observation of a non-uniform current profile in a 3D material. 
Moreover, this is a high carrier-density material in which the direct Coulomb mediated electron-electron scattering is expected to be screened, making the observation of hydrodynamic behavior surprising.
Importantly, we find that the hydrodynamic character of the current depends strongly and non-monotonically on temperature.
From an independent first principles investigation, which includes interactions that have so far been overlooked in the search for hydrodynamic candidates, namely phonon-mediated electron-electron scattering~\cite{coulter2018microscopic,levchenko_transport_2020}, we show that unlike Coulomb scattering, this dictates electron dynamics in WTe$_2$.
Using these first principles scattering rates as input to the spatially resolved Boltzmann transport equation, we predict hydrodynamic current density profiles which capture the non-monotonic temperature dependence.
The combination of experiment and theory enables us to extract the characteristic length scale of electron-electron interactions in WTe$_2$, providing strong evidence for phonon-mediated hydrodynamics. 

$Td$-WTe$_2$ crystallizes in the orthorhombic lattice (space group $Pmn2_1$), with two layers in the unit cell bonded by weak van der Waals interactions along the crystallographic $\hat{c}$ axis, shown in ~Fig.~1a. 
The semimetal is charge-compensated~\cite{zhu_quantum_2015}, as illustrated in Fig.~1b by the WTe$_2$ Fermi surface with the electron (hole) pockets displayed in blue (red).
We exfoliate a WTe$_2$ flake of approximately $60~\mathrm{nm}$ thickness, cleaved along its \(\hat{a}\) crystallographic axis. 
To image the current flow in WTe$_2$, we use a scanning probe based on a Nitrogen-Vacancy (NV) defect in diamond~\cite{maletinsky_robust_2012,pelliccione_scanned_2016,zhou_scanning_2017,xie_crystallographic_2018}, an atomic-size quantum magnetometer.
The NV is sensitive to the magnetic field parallel to its crystal axis through the Zeeman effect, and our experiments take advantage of the long coherence time of the NV to perform echo magnetometry and achieve approximately $10~\mathrm{nT}$ magnetic sensitivity (see Methods and Supplementary Sections 1, 2).
The unique custom-built cryostat and scanning system used in this study, allowing imaging at variable temperature, is shown schematically in Fig.~1c. 
Fig.~1d shows a closeup of the diamond tip with the NV defect above the WTe$_2$ sample.
At each tip location, the NV sensor detects the magnetic field generated by the current flowing along the sample. 
In our notation, the current is flowing along the $\hat{y}$-axis, generating non-zero magnetic field in the $\hat{x}$-$\hat{z}$ plane. 
The NV defect axis is oriented along the $\hat{y}$-$\hat{z}$ plane, and thus we are only sensitive to the $\hat{z}$ component of the magnetic field. 

\begin{figure*}
\centering
\includegraphics[width=\textwidth]{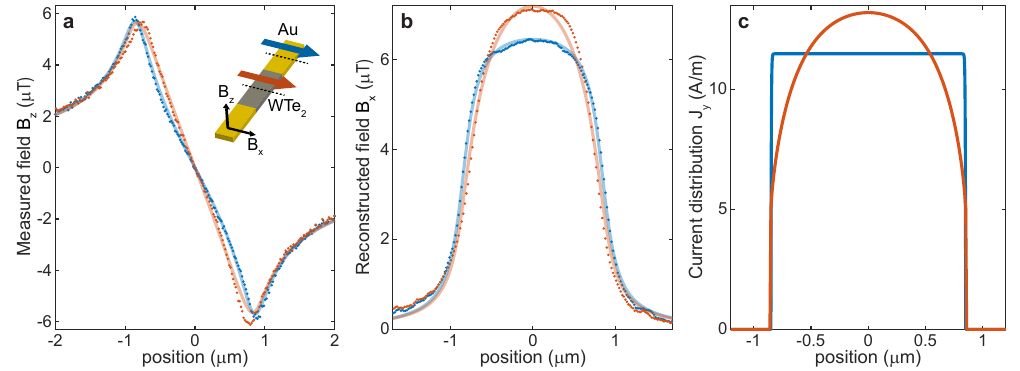}
\caption{ \textbf{Comparison of current flow profiles in WTe$_2$ and gold.}
(\textbf{a}) $B_z$ magnetic field, measured by the NV sensor in a 1D scan across a channel of width $W=1.7~\mathrm{\mu m}$. 
The blue and orange markers show measurements along the gold contact and the WTe$_2$ device respectively, normalized to have a total current $I_{tot}=20~\mathrm{\mu A}$.
The blue and orange lines correspond to the expected $B_z$ field generated by the currents in (c) at a height of $h=140~\mathrm{nm}$ measured between the NV and the middle of the WTe$_2$ flake. 
The WTe$_2$ and gold data were taken at 20 K and 90 K respectively, with the gold measurements showing negligible temperature dependence (see Supplementary Section 3).
(\textbf{b}) Reconstruction of the $B_x$ field along the scan, from the data shown in (a) for the corresponding blue and orange markers. 
The blue and orange lines correspond to the expected $B_x$ field generated by the currents in (c). 
Note that there is an asymmetry at the edges of the sample, with both curves appearing above the expected values on the left and below them on the right. 
This is likely due to a $1^{\circ}$ angle mismatch of our NV sensor.
The lack of mirror symmetry pronounced in the middle of the channel likely reflects the true current distribution. 
The source of this asymmetry is currently unknown.
(\textbf{c}) Theoretical current distributions along a channel of width $W=1.7~\mathrm{\mu m}$ with total current $I_{tot}=20~\mathrm{\mu A}$. 
The blue line shows a uniform profile, expected for diffusive flow. 
The orange line shows a curved current profile with flow enhanced in the center of the channel and reduced along the edges (see discussion of Fig.~4 for more details on the profile).}
\end{figure*}

In Fig.~2a, we present the $\hat{z}$-component of the magnetic field ($B_z$) measured by our NV tip scanning along two 1D line scans along the $\hat{x}$-axis, one taken above a gold contact (blue markers) and the other taken above our WTe$_2$ sample (orange markers).
Both measurements correspond to a channel width of $W=1.7~\mathrm{\mu m}$ and total current $I_{tot}=20~\mathrm{\mu A}$, and were taken at a tip height of $h=~140~\mathrm{nm}$ measured between the NV and the middle of the WTe$_2$ flake (see Supplementary Section 7).
We observe noticeable differences between the two scans, with the WTe$_2$ measurement showing a sharper slope in the center of the channel and an inward shift of the extrema positions. 
Such differences are discernible due to the high signal-to-noise ratio of our measurement (see Supplementary Section 2).
To visualize the difference in the underlying current profile, we obtain the $\hat{x}$-component of the magnetic field ($B_x$) by Fourier reconstruction~\cite{casola_probing_2018} of the data in Fig.~2a (see Supplementary section 8). 
This is shown by the blue and orange markers in Fig.~2b respectively, with the WTe$_2$ measurement indicating enhanced current density in the center and reduced density along the edges, suggestive of hydrodynamic flow.
These observations can be made more quantitative by examining two theoretical examples of current distributions, as shown in Fig.~2c. 
The blue line corresponds to uniform flow, where electron-electron interactions play a negligible role - as expected for diffusive behavior. 
The orange line shows a non-diffusive current distribution whose quantitative details will be discussed below.
The blue and orange lines shown in Fig.~2a,b are the calculated $B_z$ and $B_x$ for the corresponding profiles in Fig.~2c. 
The good agreement with experiment confirms that current flow in gold is indeed uniform.
Notice that the finite height offset of the NV tip convolves the current distributions for both gold and WTe$_2$, resulting in a non-flat $B_x$ profile even for a fully diffusive current distribution.
By contrast, the flow in WTe$_2$ deviates significantly from diffusive flow, even after taking the finite height into account.
We then apply our apparatus to probe the temperature dependence of electron interactions in WTe$_2$ and their influence on hydrodynamic behavior.
Fig.~3a shows a contour plot of the $B_x$ magnetic field profile across our WTe$_2$ sample, taken at different temperatures.
The magnetic field at the center of the channel becomes higher with lower temperature, until it peaks at around $10-20~\mathrm{K}$, and then becomes lower again. 
Similarly the width of the profile, highlighted by the gray contour at $B_x=4~\mathrm{\mu T}$, is also narrowest around $10-20~\mathrm{K}$.
This non-monotonic behavior~\cite{krishna_kumar_superballistic_2017} is evident in three field profiles taken at $90~\mathrm{K}$, $20~\mathrm{K}$, and $5~\mathrm{K}$ (Fig.~3b).
The profile at $20~\mathrm{K}$ is both the narrowest and has the highest peak value, indicating hydrodynamic effects on current flow are maximal at this temperature.

\begin{figure*}
\centering
\includegraphics[width=\textwidth]{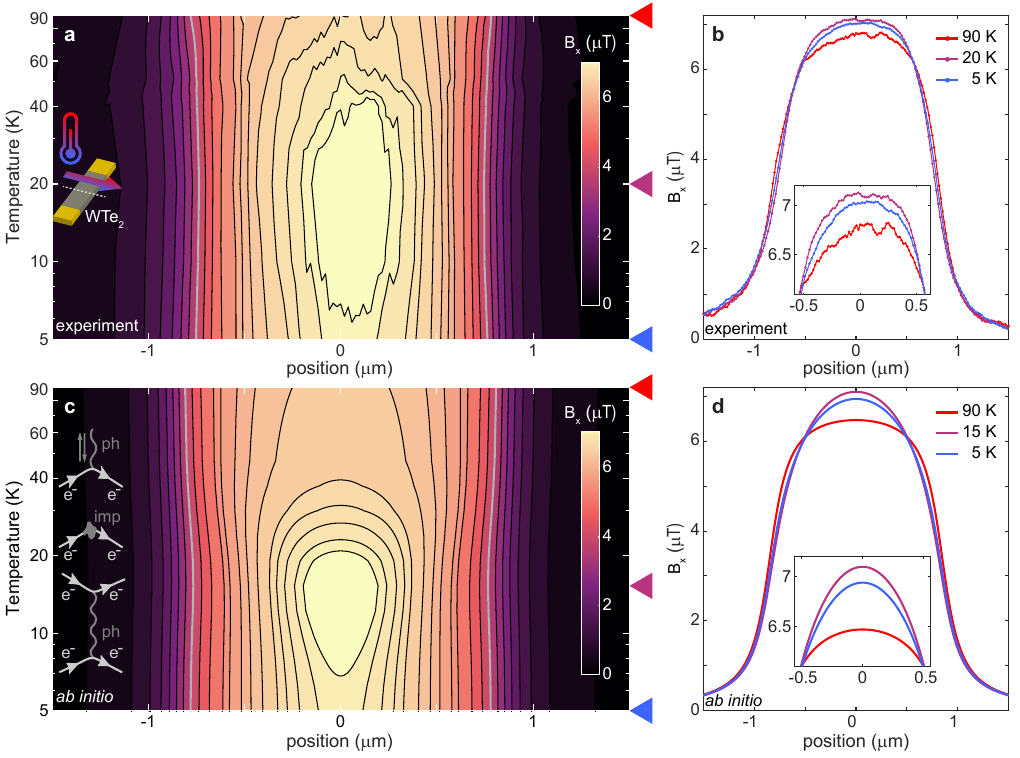}
\caption{ \textbf{Temperature-dependent magnetic field profiles.}
(\textbf{a}) Contour plot of the magnetic field profile $B_x$ across the WTe$_2$ device ($\hat{x}$-axis) at different temperatures ($\hat{y}$-axis).
The scans were taken along the same $y$-position to isolate the effects of temperature on current flow (see Supplementary Section 4).
The contour at $B_x=4~\mathrm{\mu T}$ is highlighted in gray to show the narrowest profile appears between $10-20~\mathrm{K}$, where the peak height is also maximal. 
(\textbf{b}) 3 line-cuts of (a) taken at $90~\mathrm{K}$, $20~\mathrm{K}$, and $5~\mathrm{K}$ as can be seen by the respective arrows. 
The inset zooms in on the peak of the profiles. 
The profile at $20~\mathrm{K}$, also shown in orange in Fig.~2c, is both the narrowest and has the highest peak value.
(\textbf{c}) Contour plot of the magnetic field profiles obtained from numerical transport calculations using \textit{ab initio} scattering rate inputs for electron-electron and electron-phonon interactions, and assuming an electron-impurity mean free path of $1.9~\mathrm{\mu m}$.
The theory captures the non-monotonic temperature dependence, peaking around $15~\mathrm{K}$.
(\textbf{d}) 3 line-cuts of (c) taken at $90~\mathrm{K}$, $15~\mathrm{K}$, and $5~\mathrm{K}$, showing quantitative agreement with the experimental data in (b).}
\end{figure*}

To understand the underlying microscopic origin of hydrodynamic behavior in WTe$_2$ and its non-monotonic temperature dependence, we investigate the competition of electron-electron interactions with boundary scattering and momentum-relaxing scattering against lattice vibrations and impurities in the sample.
From an independent \textit{ab initio} theory we extract the temperature dependence of the different scattering mechanisms in WTe$_2$.
These are used to solve the Boltzmann transport equation in the relaxation time approximation~\cite{de_jong_hydrodynamic_1995,moll_evidence_2016}, to obtain the steady-state current profiles (see Methods), used to extract magnetic field profiles, which are shown as a function of temperature in Fig.~3c.
The theoretical predictions capture the non-monotonic temperature dependence, with the peak (approximately $15~\mathrm{K}$) showing quantitative agreement with experiment.
Note that while the results presented in the manuscript use a circular Fermi surface assumption, relaxing this assumption does not significantly change our conclusions (see Supplementary Section 9.8).

\begin{figure*}
\centering
\includegraphics[width=\textwidth]{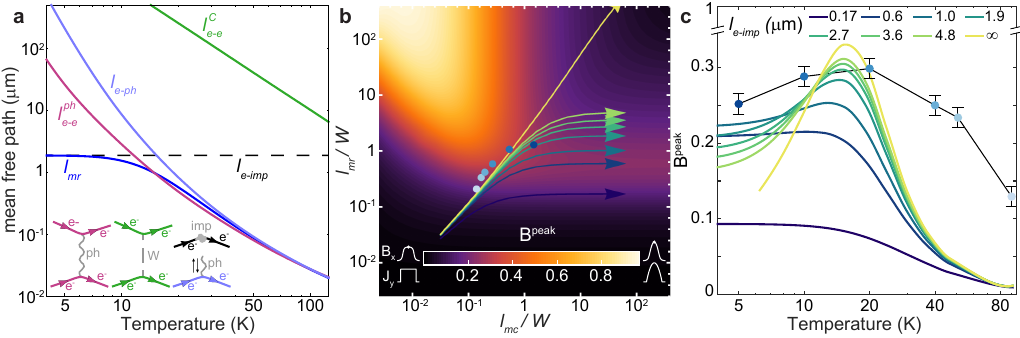}
\caption{ \textbf{\textit{ab initio} lifetimes and Boltzmann transport equation}
(\textbf{a}) Temperature dependent \textit{ab initio} electron mean free paths for WTe$_2$, obtained using density functional theory (see Methods).
We consider electron-phonon ($l_{e-ph}$), electron-impurity ($l_{e-imp}$), electron-electron scattering mediated by a screened Coulomb interaction ($l_{e-e}^C$), and by a virtual phonon ($l_{e-e}^{ph}$).
The corresponding Feynman diagrams are schematically shown in the inset.
The overall momentum-relaxing, $l_{mr}$, electron mean free path is calculated using Matthiessen's rule.
(\textbf{b}) Normalized $B_x$ magnetic field phase diagram, extracted from current profiles computed using the Boltzmann transport equation allowing for momentum-conserving ($\hat{x}$-axis) and momentum-relaxing ($\hat{y}$-axis) interactions~\cite{de_jong_hydrodynamic_1995,moll_evidence_2016}.
The color corresponds to the magnetic field $B_x$ in the center of the channel, normalized to be $0$ for uniform flow and $1$ for perfect parabolic Poiseuille flow as highlighted by the inset.
Overlaid arrows indicate the decreasing-temperature trajectories predicted \textit{ab initio} in (a) for various values of the impurity mean free path, $l_{e-imp}$, following the legend in (c).
Points correspond to fits of the magnetic field profile to the experimental data at different temperatures, following the colors in (c).
(\textbf{c}) 1D line cuts along the arrow trajectories in (b), indicating the \textit{ab initio} theory captures the non-monotonic temperature dependence of the observed hydrodynamic phenomena for $l_{e-imp}$ larger than $1 ~\mathrm{\mu m}$, after which the position of the peak at approximately $15$~K stays unchanged. 
The points correspond to experimental data at different temperatures shown in Fig.~3.
}
\end{figure*}

The theoretical current profiles can be described generally as a function of two non-dimensional parameters, the ratio of the momentum-relaxing ($l_{mr}/W$) and momentum-conserving ($l_{mc}/W$) processes' mean free paths to the channel width $W$.
Microscopically, we consider two momentum-relaxing processes, electron-phonon scattering, $l_{e-ph}$, and electron-impurity scattering, $l_{e-imp}$, and two momentum conserving processes, electron-electron scattering mediated by a screened Coulomb interaction, $l_{e-e}^C$, and by a virtual phonon, $l_{e-e}^{ph}$.
Fig.~4a shows the full temperature dependence of these mean free paths for WTe$_2$, given by our \textit{ab initio} calculation which include contributions from all electron and hole pockets (see Supplementary Section 13).
The momentum conserving mean free path is dominated by the phonon-mediated interaction, as $l_{e-e}^C$ is more than two orders of magnitude higher than $l_{e-e}^{ph}$ at all temperatures. 
In contrast, both the electron-phonon and electron-impurity scattering processes contribute to the momentum-relaxing mean free path.
We represent the resulting current profiles, as a function of these two non-dimensional parameters, by the normalized $B_x$ magnetic field peak in the middle of the channel in Fig.~4b (see Supplementary Section 9.6).
It is instructive to identify four limits in our phase-diagram: 
i) in the absence of momentum-conserving events ($l_{mc} \gg W$) and numerous momentum-relaxing events ($l_{mr} \ll W$), i.e. bottom-right corner of the phase diagram in Fig.~4b, electron flow is diffusive; 
ii) in the absence of both momentum-conserving and momentum-relaxing events ($l_{mc} \gg W$ and  $l_{mr} \gg W$, top-right), boundary effects dominate leading to ballistic flow;
iii) in the presence of numerous momentum-conserving events ($l_{mc} \ll W$) and absence of momentum-relaxing events ($l_{mr} \gg W$, top-left), electrons flow collectively and we observe hydrodynamic flow;
iv) finally, in the presence of both momentum-relaxing and momentum-conserving events ($l_{mc} \ll W$ and  $l_{mr} \ll W$, bottom-left), the flow is  momentum-`porous', and the regime is referred to as `Ziman' hydrodynamics~\cite{guyer1966Thermal}.

While Fig.~4a highlights that both $l_{mr}$ and $l_{mc}$ increase monotonically with decreasing temperature, Fig.~4b confirms there exist 1D temperature trajectories that support non-monotonic behavior.
Note that even in the absence of momentum-conserving interactions (far right of the phase diagram in Fig.~4b) current density can exhibit non-monotonic behavior, albeit with quantitative disagreement with our experiment and \emph{ab initio} calculation (see Supplementary Section 9.5).
Overlaid arrows in Fig.~4b mark the decreasing temperature trajectories following the \textit{ab initio} calculations in Fig.~4a, for empirical values of $l_{e-imp}$ between $0.17 \mathrm{\mu m}$ and $4.8\mathrm{\mu m}$.
The infinite impurity arrow corresponds to momentum relaxation only due to electron-phonon interactions.
Similarly, we can fit the experimental profiles at different temperatures to one such trajectory, shown by the overlaid points in Fig.~4b.
Notice that our measurements do not comfortably fit in any of the previously-identified limits, and thus both ballistic and hydrodynamic effects contribute to the resulting current flow.
The \textit{ab initio} predictions agree well with experimental profiles, albeit the experimental data suggests a weaker temperature-dependence and, at high temperature, theory predicts a more diffusive profile than the one measured in experiment.
This deviation may be attributed to the finite thickness of the sample, position-dependent mean-free-path due to sample imperfections, or modification of the phonon spectrum by the presence of the quartz substrate (see Supplementary Section 12).
Furthermore, relaxing the circular Fermi surface assumption, results in modified current density curvature which could also account for some of the discrepancy (see Supplementary Section 9.8).
The non-monotonic behavior is shown more clearly in Fig.~4c, which plots these 1D trajectories for empirical values of $l_{e-imp}$ between $0.17 \mathrm{\mu m}$ and $4.8\mathrm{\mu m}$ (Fig.~4c). 
While a minimum impurity mean free path is necessary to observe non-monotonic behavior, the location of the peak is largely independent of $l_{e-imp}$ above this threshold.
This allows us to extract a robust estimate for the electron-electron interaction length, of $l_{e-e}^{ph}\approx 200~\mathrm{nm}$ at $10-20~\mathrm{K}$.
The \textit{ab initio} calculation was performed on a fully relaxed WTe$_2$ lattice.
When instead the lattice is constrained to accommodate Weyl nodes, electron-electron interactions increase, resulting in the non-monotonic behavior peaking at approximately $8~\mathrm{K}$ (see Supplementary Section 11). 
This is in contrast with our observations, suggesting that our sample was not in the Weyl phase.

This work opens the possibility of observing hydrodynamic effects and electron interactions even in high carrier density materials where direct electron-electron interactions are screened (see Supplementary Section 14).
The comparison of theoretical predictions and experimental observations allows us to extract a robust estimate of the electron-electron interaction length in WTe$_2$, an important quantity for interacting systems otherwise challenging to extract experimentally.
Quantifying the magnitude of electron interactions in three-dimensional materials could reveal the microscopic mechanisms underlying strongly interacting systems, such as high-temperature superconductors and strange metals~\cite{hartnoll_theory_2007,davison_holographic_2014}.
While in this work we focus on one WTe$_2$ crystallographic orientation, our setup allows exploration and imaging of anisotropic hydrodynamic flows in 3D crystals. 
By studying different geometries and orientations of such anisotropic materials, future work will permit observations of non-classical fluid behavior such as steady-state vortices, coupling of the fluid stress to vorticity, and 3D Hall viscosity~\cite{varnavides2020electron}.

\section*{\textbf{Acknowledgements:}}
\noindent
The authors thank Mark Ku and Adam S. Jermyn for fruitful discussions. 
The authors also thank Pat Gumann and Rainer Stöhr for the initial design and construction of the microscope. 
T.X.Z. thanks Ronald Walsworth and Matthew Turner for diamond processing assistance.

\noindent
\textbf{Funding:}
This work was primarily supported by ARO Grant No. W911NF-17-1-0023 and the Gordon and Betty Moore Foundations EPiQS Initiative through Grant No. GBMF4531. Fabrication of samples was supported by the U.S. Department of Energy, Basic Energy Sciences Office, Division of Materials Sciences and Engineering under award DE-SC0019300. A.Y. also acknowledges support from ARO grants W911NF-18-1-0316, and W911NF-1-81-0206, the NSF grants DMR-1708688 and the STC Center for Integrated Quantum Materials, NSF Grant No. DMR-1231319, and the Aspen Center of Physics supported by NSF grant PHY-1607611.
G.V., Y.W., and P.N. acknowledge support from the Army Research Office MURI (Ab-Initio Solid-State Quantum Materials) Grant No. W911NF-18-1-0431 that supported development of computational methods to describe microscopic, temperature-dependent dynamics in low-dimensional materials. G.V. acknowledges support from the Office of Naval Research grant on High-Tc Superconductivity at Oxide-Chalcogenide Interfaces (Grant Number N00014-18-1-2691) that supported theoretical and computational methods for phonon-mediated interactions.
Y.W. is partially supported by the STC Center for Integrated Quantum Materials, NSF Grant No. DMR-1231319 for development of computational methods for topological materials.
This research used resources of the National Energy Research Scientific Computing Center, a DOE Office of Science User Facility supported by
the Office of Science of the U.S. Department of Energy under Contract No. DE-AC02-05CH11231 as well as resources at the Research Computing 
Group at Harvard University.
This work was performed, in part, at the Center for Nanoscale Systems (CNS), a member of the National Nanotechnology Infrastructure Network, which is supported by the NSF under award no. ECS-0335765. CNS is part of Harvard University.
P.N. is a Moore Inventor Fellow and gratefully acknowledges support through Grant No. GBMF8048 from the Gordon and Betty Moore Foundation.
C.A.C.G. acknowledges support from the NSF Graduate Research Fellowship Program under Grant No. DGE-1745303.
A. T. P. acknowledges support from the Department of Defense (DoD) through the National Defense Science \& Engineering Graduate Fellowship (NDSEG) Program.

\section*{\textbf{Author contributions:}}
\noindent
J.G., P.N., and A.Y. conceived the project.
U.V., A.H., T.X.Z., and Y.D. designed and developed the experimental setup under the guidance of A.Y.; 
N.K. and J.G. grew the bulk WTe$_2$ crystals under the guidance of C.F.;
U.V, A.H., Z.Q., and A.T.P fabricated and characterized the WTe$_2$ flakes;
G.V. implemented the numerical BTE methods under the guidance of P.A. and P.N.;
G.V., Y.W., and C.A.C.G. developed and implemented the \textit{ab-initio} theoretical methods under the guidance of P.N.;
U.V., A.H., G.V., Y.W., P.N., and A.Y. analyzed the data and discussed the results.
All authors contributed to the writing of the manuscript.

\section*{\textbf{Competing interests:}}
\noindent
The authors declare no conflict of interest.

\end{bibunit}

\begin{bibunit}
\section*{\textbf{Methods:}}
\noindent
\textbf{Sample details.} Single crystals of WTe$_2$ were grown by excess Te-flux. 
In the typical synthesis, pieces of W (Chempur, $99.95\%$) and Te (Alfa Aesar, $99.9999 \%$) were weighed according to the stoichiometric ratio of W$_{1.8}$Te$_{98.2}$ ($5~\mathrm{g}$) and kept in a quartz crucible. 
This crucible was vacuum sealed inside a quartz tube. 
The reaction content was heated to $1050~^\circ\mathrm{C}$ at a rate of $100~\mathrm{^\circ C/h}$ in a programmable muffle furnace. 
This temperature was maintained for $10~\mathrm{h}$ to get a homogenous molten phase, after which the temperature was lowered slowly to $750~^\circ\mathrm{C}$ at a rate of $1~\mathrm{^\circ C/h}$. 
At this temperature, excess molten flux was removed by centrifugation to obtain shiny plate-like single crystals of WTe$_2$. 
The high quality of the single crystal is evidenced by the high residual resistivity ratio (RRR) and quantum oscillations, shown in Extended Data Fig.~1.
The total carrier density and the average mobility of the bulk sample were extracted from magnetoresistance data (at $2~\mathrm{K}$) to be $1.4\times 10^{20}~\mathrm{cm^{-3}}$ and   $2.5\times 10^{5}~\mathrm{cm^{2}/Vs}$ respectively (see Extended Data Fig.~2). 
Note that the measurements were performed on a sample made from the same bulk crystal as the device used in the Main text. 
To capture the contributions of both electron and hole carrier at higher temperature we turn to Hall resistivity measurements using a two-band model. 
The extracted estimate for carrier density and mobility for holes and electrons is shown in Extended Data Fig.~3.
\\
We exfoliated WTe$_2$ flakes and searched for elongated samples which maintain constant width along a segment, allowing us to assume uniform current flow in this segment. 
The chosen flake was then transferred onto our quartz substrate. The flake was placed close to a gold line used for delivery of a radio frequency (RF) signal, necessary for manipulating the states of the NV. 
This RF signal required the use of an insulating substrate, as the conducting plane of a typical Silicon substrate would screen the RF signal and prevent NV manipulation. 
After placing the flake, we patterned and evaporated gold contacts following argon ion cleaning (without breaking vacuum). 
No lithographic process was performed to define the edges of our device.
Extended Data Fig.~4 shows an optical image of the WTe$_2$ sample used for the experiments shown in Main text Fig. 2-4. 
It is a $60~\mathrm{nm}$ thick and $100~\mathrm{\mu m}$ long flake. 
We scanned along a $1.7 ~\mathrm{\mu m}$ wide profile in a uniform segment of the flake (see dashed white line). 
The gold scan shown as a blue line in Main text Fig. 2b was taken on the gold contact shown to the right of the sample in the image. 
The evaporated gold was also $60~\mathrm{nm}$ thick to match the WTe$_2$ flake thickness.

\noindent
\textbf{Scanning probe details \& multiple pillar probe.}
Our magnetic sensing experiment is based on a diamond scanning probe with a Nitrogen vacancy (NV) defect close to its surface.
The design and fabrication of our probes is presented in detail in Ref.~\cite{zhou_scanning_2017}.
However, our cryogenic setup requires a variation of this design which we describe in this section. 
Our scanning tip is glued to a standard quartz tuning fork, and monitoring of the tuning fork frequency allows us to maintain constant height above our sample during the scan.
Extended Data Fig.~5a shows the two legs of our tuning fork, and a quartz rod glued to one of the legs.
The quartz rod is itself glued to a diamond cube $50~\mathrm{\mu m}$ thick, $50~\mathrm{\mu m}$ wide, and $125~\mathrm{\mu m}$ long.
A closeup of the diamond is shown in Extended Data Fig.~5b, and the $3~\mathrm{\mu m}$ etched diamond pillar can be seen. 
Our NV defect is located approximately $20~\mathrm{nm}$ away from the pillar edge. 
Notice that our tuning fork is positioned horizontally, as opposed to the vertical position which is common in atomic force microscopy (AFM) systems. 
This is due to size restrictions in our setup, imposed by the short ($700~\mathrm{\mu m}$) focal length of our objective.
In the configuration described above, our diamond tip is in contact with the measured surface.
This method was problematic for certain materials (including the WTe$_2$ we measured), as the tip was damaging the surface of the material during the scan. 
In addition, the material accumulated on the tip itself, which reduced the optical contrast of the NV and hurt our measurement. 
To overcome this problem, we fabricated diamond tips with several pillars so that one pillar is used for AFM contact, while another is used for magnetic sensing. 
This technique uses the angle control available in our setup due to goniometric motors (notice the motors label $\phi$ and $\theta$ in Main text Fig. 2a), as by tuning the angle we can control which tip will be in contact and the height of the scanning tip above the sample.
In this study, the tip height was chosen as a safety precaution as the damage of a crash can be severe to both the sample and the tip.
Extended Data Fig.~6a shows a sketch of our scanning method, where one pillar is in contact with the substrate (so it does not damage the device), while another is used for scanning the sample. 
The angle $\theta$ of the diamond tip is used to tune the height of the NV above the sample. 
Extended Data Fig.~6b shows an optical image of such a multiple pillar diamond with 7 pillars in a row. 
The outer contact pillars are made to be thicker (approximately $800~\mathrm{nm}$ diameter on the edge) so they are optimized for durability on contact. 
The central pillars are thin ($300-400~\mathrm{nm}$ diameter in the middle) so they contain one NV defect per pillar on average, and for optimal optical waveguiding.

\noindent
\textbf{Echo magnetometry measurements.}
Our magnetic field data is measured using an echo magnetometry sequence on the NV for $\tau=21~\mathrm{\mu s}$. A current of $19~\mathrm{\mu A}$ is applied through the device during this sequence, alternating in direction between the two halves of the sequence. 
To extract the phase, we repeat this experiment with four different final $\pi/2$ pulse options: $\pi/2$ pulse around X-axis, $\pi/2$ pulse around the Y-axis, $-\pi/2$ pulse around the X-axis, and $-\pi/2$ pulse around the Y-axis. 
The phase can the be extracted from the measurements by:
\begin{equation}
\varphi = \arctan \left(\frac{M_{X\pi/2} - M_{X -\pi/2}}{M_{Y -\pi/2} - M_{Y \pi/2}}\right),
\end{equation}
where $M_{X\pi/2}$ is the measurement following the $\pi/2$ pulse around X-axis and respectively for the others. For more information see Supplementary Section 1.

\noindent
\textbf{Ab initio calculations of different scattering mechanisms.}
WTe$_2$ is a semimetal with considerable density of states at the Fermi level, where the electrons mainly scatter against other electrons and phonons.
Here we consider four microscopic electron scattering events in WTe$_2$: the momentum conserving electron-electron (e-e) scattering mediated by Coulomb screening ($\tau_{\rm ee}^{\rm W}$) and by a phonon ($\tau_{\rm ee}^{\rm PH}$), plus the momentum relaxing electron-phonon (e-ph) scattering ($\tau_{\rm eph}$) and electron-impurity (e-imp) scattering ($\tau_{\mathrm{imp}}$) (Fig.~4, main text).
Details of the methodology can be found in earlier work~\cite{brown2016nonradiative,narang2016cubic,garcia2020optoelectronic,coulter2018microscopic,ciccarino2018dynamics} and a similar study on type-II Weyl semimetal WP$_2$ has successfully revealed the electron scattering microscopics~\cite{coulter2018microscopic}.

\noindent
\textit{Electron-phonon scattering.} 
For an electron with energy $\varepsilon_{n{\bf k}}$ (with $n$ band index and $\bf{k}$ wavevector) scattered by a phonon of energy $\omega_{{\bf q}\nu}$ (with ${\bf q}$ the momentum and $\nu$ the branch index), the electron final state has momentum of $\textbf{k+q}$ and energy of $\varepsilon_{m\bf{k+q}}$ (with $m$ the new band index). 
The electron-phonon scattering time ($\tau_{\rm eph}$) can be obtained from the electron self energy by Fermi's golden rule following
\begin{align}
     \label{Imsigma_eph}
       \tau^{-1}_\mathrm{eph}(n{\bf k})
       &=\frac{2\pi}{\hbar}\sum_{m\nu}\int_{\mathrm{BZ}}\frac{d\textbf{q}}{\Omega_{\mathrm{BZ}}}\left|g_{mn,\nu}(\textbf{k,q}) \right|^2 \nonumber \\
       &\quad \times [(n_{\textbf{q}\nu}+\frac{1}{2}\mp\frac{1}{2})\delta(\varepsilon_{n{\bf k}}\mp\omega_{{\bf k}\nu}-\varepsilon_{m\textbf{k}+\textbf{q}})],
\end{align}
where $\Omega_{\mathrm{BZ}}$ is the Brillouin zone volume, $f_{n\textbf{k}}$ and $n_{\textbf{q}\nu}$ are the Fermi-Dirac and Bose-Einstein distribution functions, respectively. 
The e-ph matrix element for a scattering vertex is given by 
\begin{equation}
   g_{mn,\nu}(\textbf{k,q})=\left(\frac{\hbar}{2m_0\omega_{\textbf{q}\nu}}\right)^{1/2}\langle \psi_{m{\bf k+q}} |
\partial_{{\bf q}\nu}V | \psi_{n{\bf k}}\rangle.
\end{equation}
Here $\langle \psi_{m{\bf k+q}}|$ and $|\psi_{n{\bf k}}\rangle$ are Bloch eigenstates and $\partial_{{\bf q}\nu}V$ is the perturbation of the self-consistent potential with respect to ion displacement associated with a phonon branch with frequency $\omega_{\textbf{q}\nu}$. 
The momentum-relaxing electron scattering rates are evaluated by accounting for the change in momentum between final and initial states based on their relative scattering angle following 
\begin{align}
\left(\tau_{\mathrm{eph}}^{\mathrm{mr}}(n\textbf{k})\right)^{-1}&=  \frac{2\pi}{\hbar}\sum_{m\nu}\int_{\mathrm{BZ}}\frac{d\textbf{q}}{\Omega_{\mathrm{BZ}}}\left|g_{mn,\nu}(\textbf{k,q}) \right|^2\nonumber \\ 
&\quad \times [(n_{\textbf{q}\nu}+\frac{1}{2}\mp\frac{1}{2})\delta(\varepsilon_{n\textbf{k}}\mp\omega_{\textbf{q}\nu}-\varepsilon_{m\textbf{k}+\textbf{q}})] \nonumber \\
&\quad \times \left(1-\frac{v_{n\textbf{k}}\cdot v_{n\textbf{k}}}{|v_{n\textbf{k}}| |v_{n\textbf{k}}|}\right),
\end{align}
where $v_{n\textbf{k}}$ is the group velocity.
We calculate the temperature dependent momentum relaxing {$\tau_{\rm mr}$} by taking a Fermi-surface average weighted by $|v_{n\textbf{k}}|^2$ and the energy derivative of the Fermi occupation for transport properties following 
\begin{equation}
    \tau_{\mathrm{eph}}^{\mathrm{mr}}=\frac{\int_{\mathrm{BZ}}\frac{\mathrm{d}\textbf{k}}{(2\pi)^3}\sum_n \frac{\partial f_{n\textbf{k}}}{\partial \varepsilon_{n\textbf{k}}}|v_{n\textbf{k}}|^2\tau_{\mathrm{eph}}^{\mathrm{mr}}({n\textbf{k}})}{\int_{\mathrm{BZ}}\frac{\mathrm{d}\textbf{k}}{(2\pi)^3}\sum_n\frac{\partial f_{\mathrm{n}\textbf{k}}}{\partial\varepsilon_{\mathrm{n}\textbf{k}}}|v_{n\textbf{k}}|^2}.
\end{equation}
\noindent
\textit{Phonon mediated electron-electron scattering.} 
The electron-electron scattering rate mediated by a virtual phonon can be estimated within the random phase approximation by
\begin{align}
(\tau_{ee}^{\mathrm{PH}})^{-1}&=\frac{\pi\hbar^2}{2k_BTg(\varepsilon_F)}\sum_{\nu}\int\frac{\Omega_{\mathrm{BZ}} \mathrm{d}\textbf{q}}{(2\pi)^3}G^2_{\textbf{q}\nu}\nonumber \\ 
&\quad \times \int_{-\infty}^{+\infty}\frac{\omega^2 \mathrm{d}\omega}{|\bar{\omega}_{\textbf{q}\nu}-\omega|^2\mathrm{sinh^2}\frac{\hbar\omega}{2k_BT}}.
\end{align}
Here, $g(\varepsilon_F)$ is the density of states at the Fermi level,  and $\bar{\omega}_{\textbf{q}\nu}=\omega_{\textbf{q}\nu}(1+i\pi G_{\textbf{q}\nu})$ is the complex phonon frequency corrected by the phonon-electron scattering linewidth. 
Each phonon mode is weighed within the Eliashberg spectral function by  
\begin{align}
    G_{\textbf{q}\nu}&=\sum_{mn}\int\frac{\Omega_{\rm BZ}\mathrm{d}\textbf{k}}{(2\pi)^3}|g_{mn,\nu}(\textbf{k,q})|^2\nonumber \\
    &\quad \times \delta(\varepsilon_{n\textbf{k}}-\varepsilon_F)\delta(\varepsilon_{m\textbf{k+q}}-\varepsilon_F).
\end{align}

\textit{Coulomb screening mediated electron-electron scattering.} 
The Coulomb mediated electron-electron scattering rate is obtained by the imaginary part of the quasiparticle self energy at each momentum and state ($\mathrm{Im}\Sigma_{n\textbf{k}}(n\textbf{k})$) as
\begin{align}
      \label{Imsigma_ee}
       \tau^{-1}_\mathrm{ee}(n\textbf{k})&=
       \frac{2\pi}{\hbar}\int_{\mathrm{BZ}}\frac{d\textbf{k}'}{(2\pi)^3}\sum_{n'}\sum_{\textbf{GG}'}\tilde{\rho}_{n'\textbf{k}',n\textbf{k}}(\textbf{G})
       \tilde{\rho}^*_{n'\textbf{k}',n\textbf{k}}(\textbf{G}') \nonumber \\ 
       &\quad \times\frac{4\pi e^2}{|\textbf{k}'-\textbf{k}+\textbf{G}|^2}\mathrm{Im}[\epsilon^{-1}_{\textbf{GG}'}(\textbf{k}'-\textbf{k},\varepsilon_{n,\textbf{k}}-\varepsilon_{n'\textbf{k}'})],
\end{align}
where $\tilde{\rho}_{n'\textbf{k}',n\textbf{k}}(\textbf{G})$ is the plane wave expansion of the product density $\sum_{\sigma}u^{\sigma*}_{n'\textbf{k}'(\textbf{r})}u^{\sigma}_{n\textbf{k}}(\textbf{r})   $ of the Bloch functions with reciprocal lattice vectors \textbf{G}, and $\epsilon^{-1}_{\textbf{GG}'}(\textbf{k}'-\textbf{k},\varepsilon_{n,\textbf{k}}-\varepsilon_{n'\textbf{k}'})$ is the microscopic dielectric function in a plane wave basis calculated within the random phase approximation.

We then utilize the analytical relation of $\tau_{\rm ee}$ with dependence on temperature according to the conventional Fermi-liquid theory since WTe$_2$ has a considerable density of states at $\varepsilon_F$. 
There, the electron-electron scattering rate grows quadratically away from the Fermi energy and with temperature as
\begin{equation}
    \tau^{-1}_{\mathrm{ee}}(\varepsilon,T)\approx \frac{D_e}{\hbar}[(\varepsilon-\varepsilon_F)^2+(\pi k_B T)^2].
\end{equation}
We obtain $\tau^{-1}_{\mathrm{ee}}$ by fitting all the self energies in the entire Brillouin zone for all energy bands at 298~K, extracting $D_e$ and then adding the temperature dependence~\cite{brown2016nonradiative}.

Finally, taking into account the impurity scattering, the overall momentum relaxing mean free path ($l_{mr}$) and momentum conserving mean free path ($l_{mc}$) are estimated by Matthiessen’s rule,
\begin{align}
l_{mr}&=\frac{v_F}{(\tau^{\mathrm{mr}}_{\mathrm{eph}})^{-1}+(\tau_{\mathrm{imp}})^{-1}}\nonumber \\
l_{mc}&=\frac{v_F}{(\tau^{\mathrm{PH}}_{\mathrm{ee}})^{-1}+(\tau^{\mathrm{W}}_{\mathrm{ee}})^{-1}},
\end{align}
with $v_F$ being the Fermi surface averaged velocity and $\tau_{\mathrm{imp}}$ the impurity scattering time that does not have temperature dependence but varies in different samples.

\noindent
\textbf{Boltzmann transport equation calculations.}
We solve the electronic Boltzmann transport equation numerically, in the presence of momentum-conserving scattering. 
At steady-state, the evolution of the distribution function $f(\boldsymbol{r},\boldsymbol{k})$ for non-equilibrium electrons in the neighborhood of position $\boldsymbol{r}$ with wave vector $\boldsymbol{k}$ is given by:
\begin{align}
    \boldsymbol{v}_{\boldsymbol{k}} \cdot \nabla_{\boldsymbol{r}} f(\boldsymbol{r},\boldsymbol{k}) + e \boldsymbol{E} \cdot \nabla_{\boldsymbol{k}} f(\boldsymbol{r},\boldsymbol{k}) = \Gamma \left[f\right],
\label{eq:BTE}
\end{align}
where $\boldsymbol{v}_{\boldsymbol{k}}$ is the electron group velocity, and $\Gamma\left[f\right]$ is the collision integral.
We consider flow through a 2D channel ($\boldsymbol{r}=(x,y)$) of width $W$, for a circular Fermi surface ($\boldsymbol{v}=v_{\mathrm{F}}(\cos{\theta}, \sin{\theta})$), at the relaxation time approximation level.
Under these approximations, we linearize~\cref{eq:BTE}, to give an integro-differential equation in terms of an `effective' mean free path, $l_{\mathrm{eff}}$~\cite{de_jong_hydrodynamic_1995}:
\begin{align}
   \sin(\theta) \partial_y \, {l}_{\mathrm{eff}}(y,\theta) + \frac{{l}_{\mathrm{eff}}(y,\theta)}{l} = 1 + \frac{\tilde{l}_{\mathrm{eff}}}{l_{\mathrm{mc}}},
\label{eq:BTE-hydro}
\end{align}
where we use Mathhiessen's rule $l^{-1} = l_{mc}^{-1} + l_{mr}^{-1}$ to combine the depopulation of electronic states with a combined mean free path $l$.
Note our treatment implicitly assumes a single electronic carrier, neglecting transport effects due to electron-hole collisions~\cite{Nam2017}.
We justify this assumption by computing the relative contributions of scattering between different charge carriers, and show that they can be safely ignored (see Supplementary Section 13).
Momentum-conservation is accounted for by the last term, which defines the `average' mean free path $\tilde{l}_{\mathrm{eff}}(y)$ and is directly proportional to current density, $j_x(y)$:
\begin{align}
    \tilde{l}_{\mathrm{eff}}(y) &= \int_0^{2\pi} \frac{d \theta}{\pi} \cos^2(\theta) l_{\mathrm{eff}}(y,\theta) \\
    j_x(y) &= \left(\frac{m}{\pi \hbar^2}\right) \varepsilon_{\mathrm{F}} e^2 \frac{E_x}{m v_{\mathrm{F}}} \tilde{l}_{\mathrm{eff}}(y).
    \label{eq:BTE-4}
\end{align}
where $m$ is the electronic effective mass, $E_x$ is the electric field component in the $x$ direction, and $v_F$ is the Fermi velocity. 
Equation~\cref{eq:BTE-hydro} is solved numerically by transforming it into a Fredholm integral equation of the 2\textsuperscript{nd} kind~\cite{de_jong_hydrodynamic_1995}:
\begin{align}
    \tilde{l}_{eff}(y) - \frac{1}{l_{mc}} &\int_{-W/2}^{W/2} dy' K(y,y') \tilde{l}_{eff}(y') = \tilde{l}^{(0)}_{eff}(y) \\
    \tilde{l}^{(0)}_{eff}(y) = l - \frac{2l}{\pi} &\int_0^{\pi/2} d\phi \cos^2(\phi) \\
    &\times \left\{{\rm exp}\left(-\frac{W/2+y}{l \sin(\phi)}\right) + {\rm exp}\left(-\frac{W/2-y}{l\sin(\phi)}\right)\right\} \nonumber \\
    K(y,y') := \frac{2}{\pi}&\int_0^{\pi/2} d\phi \frac{\cos^2(\phi)}{\sin(\phi)} {\rm exp}\left(-\frac{\left| y-y'\right|}{l \sin(\phi)}\right)
\end{align}
To ensure numerical stability across a wide range of $l_{mr(mc)}/W$, we solve this using the modified quadrature method~\cite{rahbar2008computational}.

\section*{\textbf{Data availability}}
\noindent
Data for figures that support the manuscript are available at: \url{https://doi.org/10.7910/DVN/ULRIZG}.
All other data that support the findings of this paper are available from the authors upon request.

\section*{\textbf{Code availability}}
\noindent
Source codes for the \textit{ab initio} model and Boltzmann transport equation calculation are available from the authors upon request



%

\end{bibunit}

\end{document}


\title{Supplementary material for \\\textit{Imaging phonon-mediated hydrodynamic flow in WTe$_2$}} 

\author{Uri Vool}
\thanks{These authors contributed equally.}
\affiliation{John Harvard Distinguished Science Fellows Program, Harvard University, Cambridge, MA 02138, USA}
\affiliation{Department of Physics, Harvard University, Cambridge, MA 02138, USA}
\author{Assaf Hamo}
\thanks{These authors contributed equally.}
\affiliation{Department of Physics, Harvard University, Cambridge, MA 02138, USA}
\author{Georgios Varnavides}
\thanks{These authors contributed equally.}
\affiliation{John A. Paulson School of Engineering and Applied Sciences, Harvard University, Cambridge, MA 02138, USA}
\affiliation{Department of Materials Science and Engineering, Massachusetts Institute of Technology, Cambridge, MA 02139, USA}
\affiliation{Research Laboratory of Electronics, Massachusetts Institute of Technology, Cambridge, MA 02139, USA}
\author{Yaxian Wang}
\thanks{These authors contributed equally.}
\affiliation{John A. Paulson School of Engineering and Applied Sciences, Harvard University, Cambridge, MA 02138, USA}
\author{Tony X. Zhou}
\affiliation{Department of Physics, Harvard University, Cambridge, MA 02138, USA}
\affiliation{John A. Paulson School of Engineering and Applied Sciences, Harvard University, Cambridge, MA 02138, USA}
\affiliation{Research Laboratory of Electronics, Massachusetts Institute of Technology, Cambridge, MA 02139, USA}
\author{Nitesh Kumar}
\affiliation{Max-Planck-Institut für Chemische Physik Fester Stoffe, Dresden, Germany}
\author{Yuliya Dovzhenko}
\affiliation{Department of Physics, Harvard University, Cambridge, MA 02138, USA}
\author{Ziwei Qiu}
\affiliation{Department of Physics, Harvard University, Cambridge, MA 02138, USA}
\affiliation{John A. Paulson School of Engineering and Applied Sciences, Harvard University, Cambridge, MA 02138, USA}
\author{Christina A. C. Garcia}
\affiliation{John A. Paulson School of Engineering and Applied Sciences, Harvard University, Cambridge, MA 02138, USA}
\author{Andrew T. Pierce}
\affiliation{Department of Physics, Harvard University, Cambridge, MA 02138, USA}
\author{Johannes Gooth}
\affiliation{Department of Physics, Harvard University, Cambridge, MA 02138, USA}
\affiliation{Max-Planck-Institut für Chemische Physik Fester Stoffe, Dresden, Germany}
\affiliation{Institut für Festkörper- und Materialphysik, Technische Universität Dresden, 01062 Dresden, Germany}
\author{Polina Anikeeva}
\affiliation{Department of Materials Science and Engineering, Massachusetts Institute of Technology, Cambridge, MA 02139, USA}
\affiliation{Research Laboratory of Electronics, Massachusetts Institute of Technology, Cambridge, MA 02139, USA}
\author{Claudia Felser}
\affiliation{John A. Paulson School of Engineering and Applied Sciences, Harvard University, Cambridge, MA 02138, USA}
\affiliation{Max-Planck-Institut für Chemische Physik Fester Stoffe, Dresden, Germany}
\author{Prineha Narang}
\email[Electronic address:\;]{prineha@seas.harvard.edu}
\affiliation{John A. Paulson School of Engineering and Applied Sciences, Harvard University, Cambridge, MA 02138, USA}
\author{Amir Yacoby}
\email[Electronic address:\;]{yacoby@g.harvard.edu}
\affiliation{Department of Physics, Harvard University, Cambridge, MA 02138, USA}
\affiliation{John A. Paulson School of Engineering and Applied Sciences, Harvard University, Cambridge, MA 02138, USA}
\date{\today}


\baselineskip24pt

\maketitle 

\renewcommand\figurename{Extended Data Fig.}

\begin{figure}
\centering
\includegraphics[width=\textwidth]{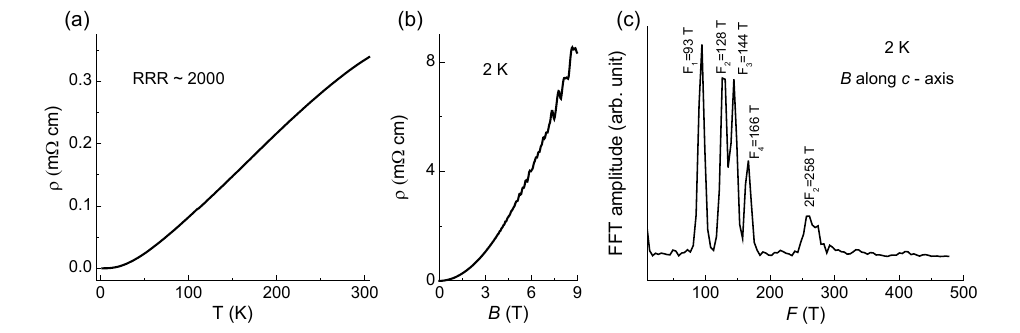}
\caption{\textbf{a} Temperature dependent resistivity at zero magnetic field. The residual resistivity ratio (RRR) between 300 K and 2 K is $\sim 2000$, signifying the high quality of the single crystal. 
\textbf{b} Magnetic field (applied along the $\hat{c}$-axis) dependent resistivity showing Shubnikov-de Haas oscillations at 2 K. 
\textbf{c} The fast Fourier transform (FFT) of the background-subtracted data of (b) when plotted against the inverse of the magnetic field. We notice four fundamental peaks corresponding to the extremal orbits of the electron and hole pockets in the Brillouin zone \cite{zhu_quantum_2015}.}
\label{fig:s_goodsample}
\end{figure}

\begin{figure}
\centering
\includegraphics[width=\textwidth]{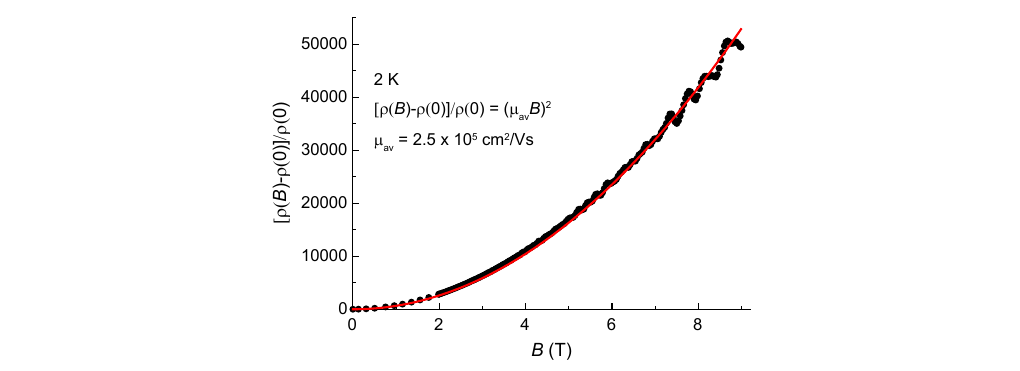}
\caption{Fitting of the magnetic field dependent resistivity at 2 K according to the relation $\frac{\rho(B)-\rho(0)}{\rho(0)}=(\mu_{av}B)^2$ where $\mu_{av}$ is the average mobility\cite{zhu_quantum_2015,ali_correlation_2015}.
This allows us to extract $\mu_{av} = 2.5\times10^5~\mathrm{cm^2/Vs}$. Using the relation $\sigma = \mu n e$, the average carrier density $n=1.47\times10^{20}~\mathrm{cm^{-3}}$ can be extracted.}
\label{fig:s_magnetores}
\end{figure}

\begin{figure}
\centering
\includegraphics[width=\textwidth]{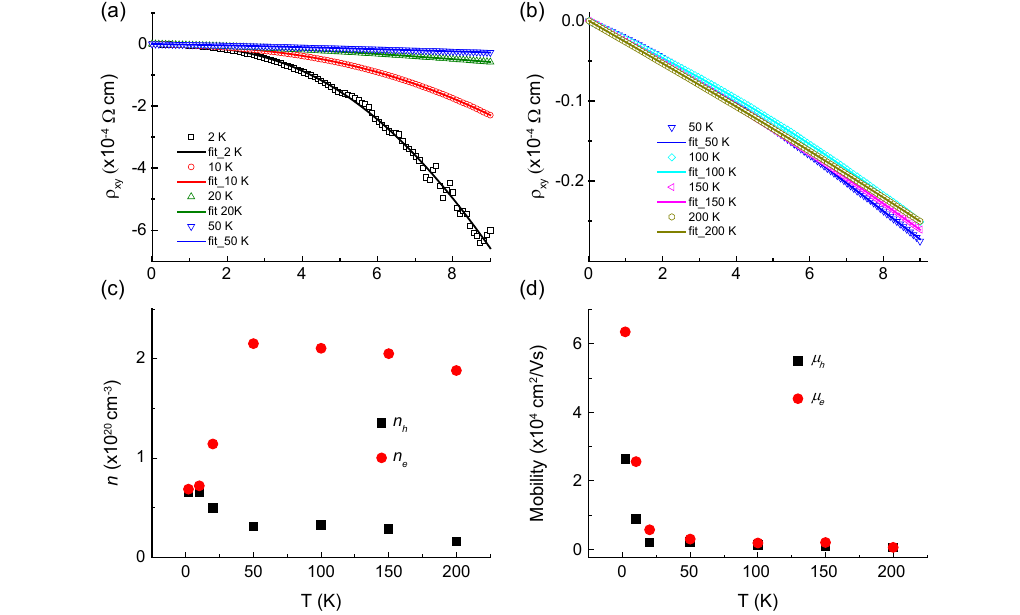}
\caption{\textbf{a}, \textbf{b} Hall resistivity vs. magnetic field measured at different temperatures. 
The data is fitted according to the two-band model $\rho_{yx} = \frac{B}{e} \frac{(n_h \mu_{h}^2-n_e \mu_{e}^2)+(n_h-n_e)\mu_{h}^2\mu_{e}^2B^2}{(n_h \mu_{h}+n_e \mu_{e})^2+(n_h-n_e)^2\mu_{h}^2\mu_{e}^2B^2}$  \cite{luo_hall_2015,PhysRevB.92.041104}. 
\textbf{c} Carrier density of electrons and holes vs. temperature, extracted from the fits in (a) and (b). 
\textbf{d} Mobility of electrons and holes vs. temperature, extracted from the fits in (a) and (b).}
\label{fig:s_carriervtemp}
\end{figure}

\begin{figure}
\centering
\includegraphics[width=\textwidth]{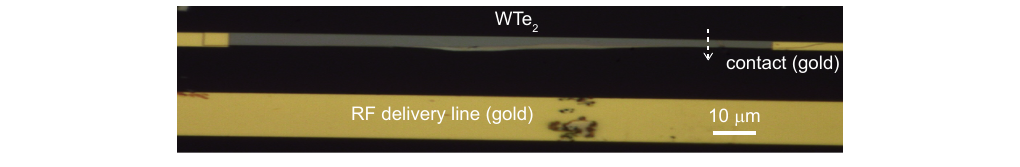}
\caption{Optical image of the WTe$_2$ sample used for the majority of the data in the manuscript, including Main text Fig. 2, 3, and 4. 
The $100~\mathrm{\mu m}$ long flake is contacted at its ends with gold contacts. 
The path along which our magnetic field profiles were taken is shown in white. 
An additional gold line nearby is used for delivery of RF power to manipulate the NV.}
\label{fig:s1}
\end{figure}

\begin{figure}
\centering
\includegraphics[width=\textwidth]{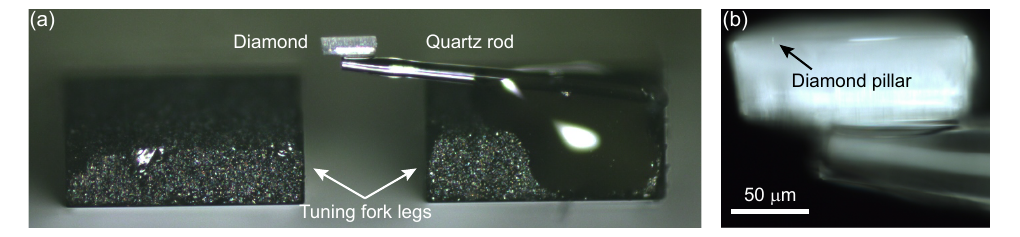}
\caption{\textbf{a} An optical image of our scanning tip. A quartz rod is glued to the edge of one of the prongs of a quartz tuning fork.
A diamond cube of $50 \times 50\times 125~\mathrm{\mu m}$ is glued to the other side of the rod.
\textbf{b} A closeup image of the diamond cube, with a $3~\mathrm{\mu m}$ etched pillar visible.
This pillar is used as our scanning probe, and an NV defect is located  $~20~\mathrm{nm}$ away from the pillar edge.}
\label{fig:s2}
\end{figure}

\begin{figure}
\centering
\includegraphics[width=\textwidth]{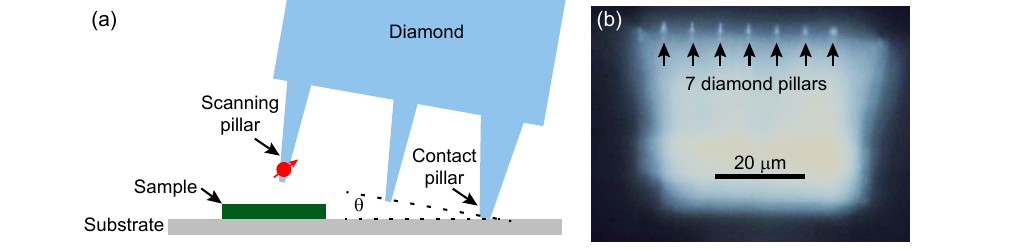}
\caption{\textbf{a} A schematic for multiple pillar sensing. 
The diamond is tilted at an angle $\theta$ (using goniometric motors) so that one pillar is in contact with the substrate, while the scanning pillar maintains a constant height above the sample without touching it. 
The angle $\theta$ control the height of the NV probe above the sample.
\textbf{b} An optical image of a multiple pillar scanning tip with 7 pillars. 
The extreme pillars are made to be thicker so they are optimized for durability in contact, while the central pillars are optimized for NV sensing.}
\label{fig:s3}
\end{figure}

\clearpage

\renewcommand\figurename{Fig.}
\renewcommand{\theequation}{S\arabic{equation}}
\renewcommand{\thefigure}{S\arabic{figure}}
\renewcommand{\theHfigure}{S\arabic{figure}}
\setcounter{figure}{0}

\section{Echo magnetometry}
There are several ways to extract the magnetic field from the NV defect, and here we will describe the method used in the paper, which utilizes the long coherence of the NV as a quantum sensor. 

The NV electron spin is a spin-1 system, and its low energy state are the triplet states $\ket{-1} = \ket{\downarrow \downarrow}$, $\ket{0} = \frac{1}{\sqrt{2}}(\ket{\uparrow \downarrow}+\ket{\downarrow \uparrow})$, and $\ket{1} = \ket{\uparrow \uparrow}$. 
The $\ket{0}$ is split in energy from $\ket{-1}$ and $\ket{1}$ by the zero field splitting and $\ket{-1}$ and $\ket{1}$ are split by applying a bias magnetic field due to Zeeman splitting. 
We apply an external bias field of $1650~\mathrm{G}$ parallel to the NV axis, where the energy splitting between the $\ket{-1}$ and $\ket{0}$ states is $1.7~\mathrm{GHz}$. 
This is the only transition used in our experiment, and so the system can now be described as a qubit made of these two states.

An external magnetic field parallel to the NV axis shifts the frequency of our qubit by $g \mu_B B_{||}$ due to the Zeeman effect, where $g \sim 2$ is the NV g-factor, $\mu_B$ is the Bohr magneton, and $B_{||}$ is the applied parallel magnetic field. This shift is approximately $2.87~\mathrm{MHz/G}$.
Measuring the resonance frequency of the qubit allows us to infer the local parallel magnetic field. 

This frequency is often measured by a continuous spectroscopy method (ESR) in which the NV is probed with RF radiation at different frequencies to find the frequency at which the NV is excited, leading to reduced red counts in the optical measurement. 
While this method is simple, it has several drawbacks. 
The continuous drive can heat up the system being measured, and lead to broadening of the resonance peak. 
Also, the contrast in this measurement is halved as the states compared are $\ket{0}$ and an equal mixture of $(\ket{0}, \ket{-1})$ as opposed to full contrast between $\ket{0}$ and $\ket{-1}$.
Another method is Ramsey interferometry, in which the qubit is placed in a superposition state $\frac{1}{\sqrt{2}}(\ket{0}+i\ket{-1})$. 
This superposition evolves to acquire a phase $\frac{1}{\sqrt{2}}(\ket{0}+ie^{i\Delta t}\ket{-1})$ due to the change $\Delta$ in the qubit resonance frequency. 
A measurement of this phase provides the shift in resonance frequency and therefore the magnetic field. 
This technique requires pulsed control, and thus overcomes many of drawbacks of continuous spectroscopy. 
The time over which we can acquire a phase is limited by the low frequency qubit coherence time $T_2^*$.

\begin{figure}
\centering
\includegraphics[width=\textwidth]{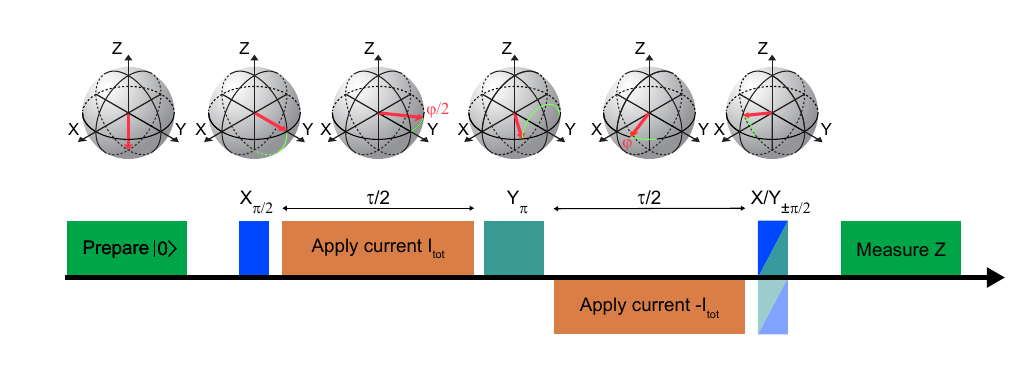}
\caption{A schematic of our echo magnetometry experiment. 
The pulses applied are shown below in sequence, and the Bloch sphere representation above shows the state of the qubit at the corresponding place in the sequence. 
The red arrow indicates the current state of the qubit, while the green line indicates the path it took on the Bloch sphere since the last step. 
A detailed description of the sequence can be found in Section 3. 
For the last $\pi/2$, there are four possible rotations, and the Bloch sphere representation shows the $\pi/2$ around the X-axis.}
\label{fig:s4}
\end{figure}

Pulsed manipulation of the qubit allows for more complicated sequences which cancel the effect of low frequency noise, improving the qubit coherence time from $T_2^*$ to $T_2$. 
In this experiment we used the simplest such sequence, the echo experiment, as shown in detail in Supplementary Fig.~\ref{fig:s4}. 
The qubit is initially prepared in the $\ket{0}$ state by illuminating green light. Afterwards, an RF pulse is applied to perform a $\pi/2$ pulse around the X-axis. 
This places the qubit in the superposition state $\frac{1}{\sqrt{2}}(\ket{0}+i\ket{-1})$. 
Then a current $I_{tot}$ is applied along the device for a duration $\tau/2$. 
This current leads to a parallel magnetic field $B_{||}$ and thus to a frequency shift $\Delta = g \mu_B B_{||}$ in the qubit resonance frequency shift. 
During a time $\tau/2$ a phase $\varphi/2 = \Delta \tau/2$ is acquired by the qubit, which is then left in the state $\frac{1}{\sqrt{2}}(\ket{0}+ie^{\varphi/2}\ket{-1})$. 
A $\pi$ pulse around the Y-axis is then applied on the qubit, leading to the state $\frac{1}{\sqrt{2}}(\ket{-1}-ie^{\varphi/2}\ket{0}) = \frac{1}{\sqrt{2}}(\ket{0}+ie^{-\varphi/2}\ket{-1})$. 
Note that the phase accumulated has been reversed and so applying the same current again will lead to a cancellation of the acquired phase. 
Instead we apply the opposite current, leading to an acquisition of an additional $-\varphi/2$ phase, leading to the state $\frac{1}{\sqrt{2}}(\ket{0}+ie^{-\varphi}\ket{-1})$. 
The phase has now been acquired and we need to extract it. 
For this, we repeat this experiment with four different pulse options: $\pi/2$ pulse around X-axis, $\pi/2$ pulse around the Y-axis, $-\pi/2$ pulse around the X-axis, and $-\pi/2$ pulse around the Y-axis. 
For each option we then measure the $Z$ component of the qubit by applying green laser light and collecting red photons. 
From the 4 measurements of probabilities to be in the $\ket{-1}$ state collected after their respective pulses we can extract $\varphi$ as:
\begin{equation}
\varphi = \arctan \left(\frac{M_{X\pi/2} - M_{X -\pi/2}}{M_{Y -\pi/2} - M_{Y \pi/2}}\right),
\end{equation}
where $M_{X\pi/2}$ is the measurement following the $\pi/2$ pulse around the X-axis and respectively for the others. 
The different measurements allow us to cancel the effect of background counts and focus only on the optical contrast due to the qubit state.

Note that this sequence is only useful for sensing alternating magnetic fields, where the field frequency is equal to the echo frequency. 
This is useful in current measurement where we can lock the application of current to the echo measurement, but is not useful for measuring static magnetic fields due to magnetization.

As mentioned above, this measurement cancels low frequency noise and thus allows us to acquire a signal up to $\tau = T_2$ as opposed to $T_2^*$. 
The probe we used had long coherence time $T_2=150~\mathrm{\mu s}$ (and $T_2^*=3~\mathrm{\mu s}$). 
However, as will be discussed in Section 6, our measurements were taken at shorter times $\tau=21~\mathrm{\mu s}$.
This was chosen because we can afford to use relatively high current, for which a longer integration time would lead to phase accumulation significantly above $2\pi$ - leading to ambiguity in the magnetic field as we can only determine the phase up to $2\pi$.

\section{Magnetic noise estimations of the NV detector}

The measurements that are shown in the main text had to be measured in exceptional signal-to-noise ratio (SNR) in order to observe the slight changes over a small temperature range. 
In order to achieve that, we made use of a combination of relatively high current $(15-20~{\mu}A)$ together with our long echo time (${\approx}20~{\mu}s$). 
In this section we will try to estimate quantitatively the SNR that we have in our measurements. 
The signal (in $\mathrm{Counts/s}$) is given by:
\begin{equation}
Signal={\Delta}{\phi}{\cdot}C{\cdot}F{\cdot}T{\cdot}N_{Avg},
\end{equation}
where ${\Delta}{\phi}$ is the accumulated phase in the echo experiment, $C$ is the contrast, $F$ is the fluorescence of the NV in $\mathrm{Counts/s}$ , $T$ is the avalanche photo-diode (APD) acquisition time, and $N_{Avg}$ is the number of averages per second. 
The noise in the system is assumed to be only shot noise:
\begin{equation}
Noise=\sqrt{F{\cdot}T{\cdot}N_{Avg}}.
\end{equation}
We want to calculate the sensitivity per second, assuming a SNR of 1:
\begin{equation}
\frac{Signal}{Noise} =1={\Delta}{\phi}{\cdot}C{\cdot}\sqrt{F{\cdot}T}{\cdot}\sqrt{N_{Avg}}
\end{equation}
From that we can extract the minimal ${\Delta}{\phi}$: 
\begin{equation}
{\Delta}{\phi}=\frac{1}{C{\cdot}\sqrt{F{\cdot}T}}{\cdot}\frac{1}{\sqrt{N_{Avg}}}
\end{equation}
In order to estimate the smallest magnetic field, we need to convert phase to field via:
\begin{equation}
{\Delta}B={\Delta}{\phi}{\cdot}\frac{1}{2{\pi}{\gamma}}{\cdot}\frac{1}{{\tau}},
\end{equation}
where $\Delta B$ is the magnetic field, $\tau$ is the spin-echo time, and ${\gamma}=g{\cdot}{\mu}_B=2.87~\frac{\mathrm{MHz}}{\mathrm{Gauss}}$  where $g$ is the g-factor and ${\mu}_B$ is the permeability of the vacuum. 
From that we get:
\begin{equation}
{\Delta}B=\frac{1}{2{\pi}{\gamma}{\cdot}C\sqrt{F{\cdot}T}}{\cdot}\frac{1}{{\tau}}{\cdot}\frac{1}{\sqrt{N_{Avg}}}
\end{equation}
To estimate the magnetic noise, we can approximate (by ignoring overheads and assuming long enough ${\tau}$) $N_{Avg}{\approx}\frac{1}{{\tau}}$ (where we assume the total time spent measuring is $1~\mathrm{s}$). 
From that we get:
\begin{equation}
B_{noise}{\approx}\frac{1}{2{\pi}{\gamma}{\cdot}C\sqrt{F{\cdot}T}}{\cdot}\frac{1}{\sqrt{{\tau}}}
\end{equation}
In our NV center $C=0.15$ , $F=120{\cdot}10^3~   \mathrm{Counts/s}$ , $T=300~\mathrm{ns}$, and for our experiments in the main text ${\tau}=21.2~ \mathrm{{\mu}s}$. From these values we can estimate the magnetic field noise level:
\begin{equation}
B_{noise}=43~ \mathrm{\frac{ nT}{\sqrt{Hz}}}
\end{equation}
By comparing to theory, we empirically estimate of the noise in the magnetic field profile of the gold contact in Main text Fig. 2 and get:
\begin{equation}
B_{noise}=104~ \mathrm{\frac{ nT}{\sqrt{Hz}}}
\end{equation}
The additional noise can be attributed to mechanical noise and other overheads. 
In our measurements, in order to reduce the noise even further we were averaging over $10~\mathrm{s}$ which gives a noise of  ${\Delta}B{\approx}13~\mathrm{nT}$. 
Together with a signal of ${\sim}5~\mathrm{{\mu}T}$ (peak to peak in Main text Fig. 2b without normalization) we get an estimate of the SNR of:
\begin{equation}
\frac{Signal}{Noise}~{\approx}~384
\end{equation}
We also want to extract from this magnetic noise the minimal current density that is possible to detect using our probe. 
Assuming a simple rectangular current profile, we can use the $B_z$ profile from Main text Fig. 2b, where we applied ${\sim}13~\mathrm{{\mu}A}$ when the device width is $1.7~\mathrm{{\mu}m}$ and got ${\sim}5~\mathrm{{\mu}T}$ peak to peak. 
To achieve a SNR of $1$ for a $10~\mathrm{s}$ measurement, we get the minimal current density:
\begin{equation}
{\Delta}J~{\approx}~20 ~\mathrm{\frac{nA}{{\mu}m}}
\end{equation}
By taking advantage of our long $T_2$ and measuring for $150~\mathrm{{\mu}s}$ echo time, we are able to decrease this number even more and get:
\begin{equation}
{\Delta}J~{\approx}~7.3 ~\mathrm{\frac{nA}{{\mu}m}}
\end{equation}

\section{Cuts along the gold contact at different temperatures}

In Main text Fig.~2c, the cuts along the gold line is used as a control which shows Ohmic flow, as opposed to the WTe$_2$ current profile. In Main text Fig.~3, we show the temperature dependence of the current profile in WTe$_2$. The current profile of the gold contact is expected to be Ohmic at all temperatures. To verify this point, we measure the flow profile at 90 K and 10 K and they are shown in Supplementary Fig.~\ref{fig:sGold}. The similarity between the two curves is comparable to our noise level, unlike the WTe$_2$ profiles which show a significant deviation.

\begin{figure}
\centering
\includegraphics[width=\textwidth]{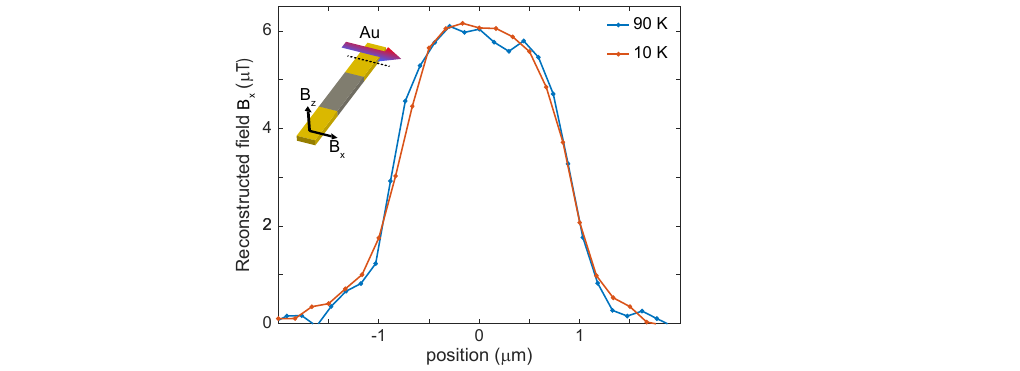}
\caption{Reconstruction of the magnetic field $B_x$ as a function of the position along the gold contact, taken at 90 K and 10 K.}
\label{fig:sGold}
\end{figure}

\section{Finding the same line-cut at different temperatures}
\begin{figure}
\centering
\includegraphics[width=\textwidth]{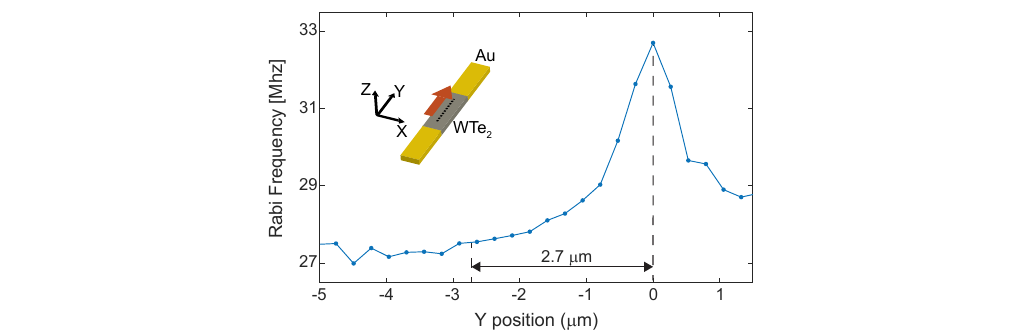}
\caption{Rabi frequency measured as a function of the position along y direction while keeping the x position along the center of the device. 
The peak is indicated by dashed line and measured $2.7~\mathrm{\mu m}$ away another dashed line indicates the position in which we performed the x scan (the scan in figure 3 of the main text).}
\label{fig:s5}
\end{figure}

In Main text Fig.~3 we have shown a dependence of the current profile on temperature. 
Due to mechanical deformations (thermal expansion) of the scanning system at different temperatures, we had to determine the exact location along the device to perform the scan for every temperature. 
We first used our optical microscope to determine the position roughly (up to a few microns).  
Following that, we performed a scan along the device (Inset, Supplementary Fig.~\ref{fig:s5}) while keeping the tip at the center of the device. 
A point along the device happened to have a strong distortion of the current which manifested as a peak in the Rabi frequency at this point. 
In Supplementary Fig.~\ref{fig:s5} the measured Rabi frequency as a function of position is shown and a clear peak is visible at around $Y=0~\mathrm{\mu m}$. 
The peak serves as a marker, and for each temperature we performed our scan at the same distance ($2.7~\mathrm{\mu m}$) from the peak (dashed vertical lines Supplementary Fig.~\ref{fig:s5}). 
Using this technique, we were able to verify the location along the device for every current profile scan in Main text Fig. 3. 

\section{Temperature dependence of current profile: additional data}
To verify the reproducibility of the non-monotonic current profile temperature dependence shown in Main text Fig.~3, we took additional cuts at different $Y$ positions along the sample. One such set of cuts is shown in Supplementary Fig.~\ref{fig:sMorecuts}. Note that these line cuts were taken without the careful $Y$ alignment procedure described in the previous section. Thus, the data shown in the Main text is more reliably free from measurement artefacts.

\begin{figure}
\centering
\includegraphics[width=\textwidth]{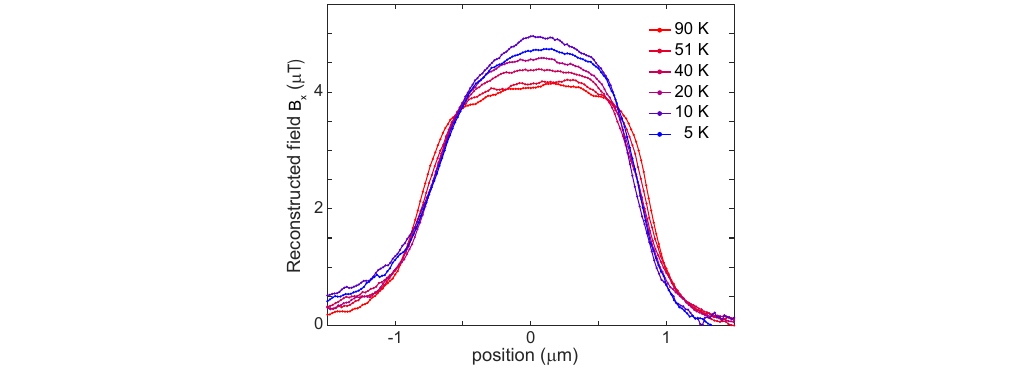}
\caption{Reconstruction of the magnetic field $B_x$ as a function of the position for various temperatures. Additional data showing non-monotonic temperature dependence.}
\label{fig:sMorecuts}
\end{figure}

\section{Flow profile dependence on current}
The bias applied on the WTe$_2$ that was used throughout the paper was chosen to be small enough that it is in the linear response regime yet large enough that the signal to noise ratio is maximized.

In order to verify linear response, we performed two scans with two different currents, one scan with relatively high current ($19~\mathrm{\mu A}$) and one scan with relatively low current ($4.7~\mathrm{\mu A}$). 
Importantly, to preserve the signal to noise ratio, we performed the two scans with different corresponding echo times to accumulate the same final phase (see Section 3 for details on AC magnetometry). 
For the high current scan, we used a $21~\mathrm{\mu s}$ phase accumulation time and for the low current scan we used $86~\mathrm{\mu s}$ (which is the same ratio between the current amplitudes). 
As a result, the low current scan was $~4$ times longer.  

\begin{figure}
\centering
\includegraphics[width=\textwidth]{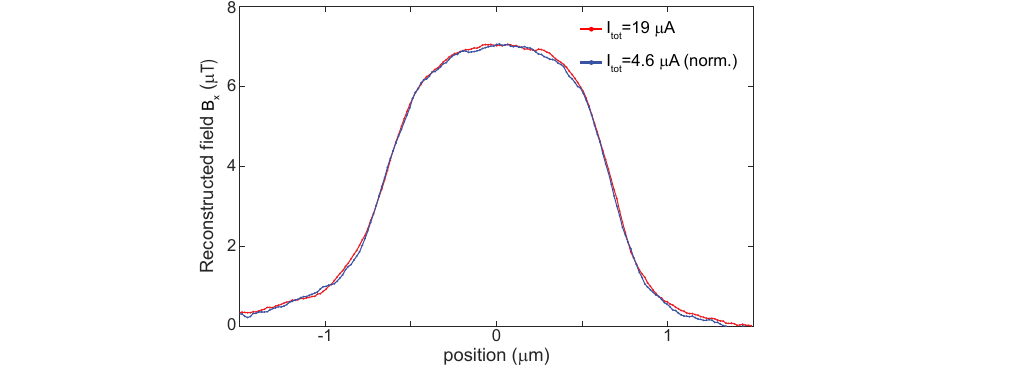}
\caption{Reconstructed magnetic field $B_x$ as a function of the position along the device, taken for two different current amplitudes. 
The red curve corresponds to a scan with a current of $19~\mathrm{\mu A}$ and measured with an echo time of $86~\mathrm{\mu s}$. 
The blue curve corresponds to a scan with current of $4.6~\mathrm{\mu A}$ and measured with an echo time of $86~\mathrm{\mu s}$. 
Both measurements were taken at $12~\mathrm{K}$.}
\label{fig:s6}
\end{figure}

The results can be seen in Supplementary Fig.~\ref{fig:s6}, as two $B_x$ magnetic field profiles are shown. 
The blue one was taken with high current and the red with low current. 
As can be clearly seen, the difference between the two curves is very small which means that indeed we are in the linear response regime. 
So, in order to maximize signal to noise ratio and to avoid drifts of the mechanical scanning system over long scanning times we used the high current $19~\mathrm{\mu A}$ and $21~\mathrm{\mu s}$ echo time throughout the paper.

\section{Magnetic field profile sample thickness dependence}
In Main Text Fig.~2a,b we compare our data with the magnetic field profiles generated by the current profiles shown in Main Text Fig.~2c at our probe height. We use the approximation of current flowing in a 2D sheet, not accounting for our sample thickness of $60~\mathrm{nm}$. 

To confirm that this is a good approximation, we show in Supplementary Fig.~\ref{fig:s_thickthin} the magnetic field profiles $B_x$ and $B_z$ generated by a uniform current profile for a 2D sheet (red line) and a $60~\mathrm{nm}$ thick sample (blue line). The good agreement between the profiles justifies our approximation. 

\begin{figure}
\centering
\includegraphics[width=\textwidth]{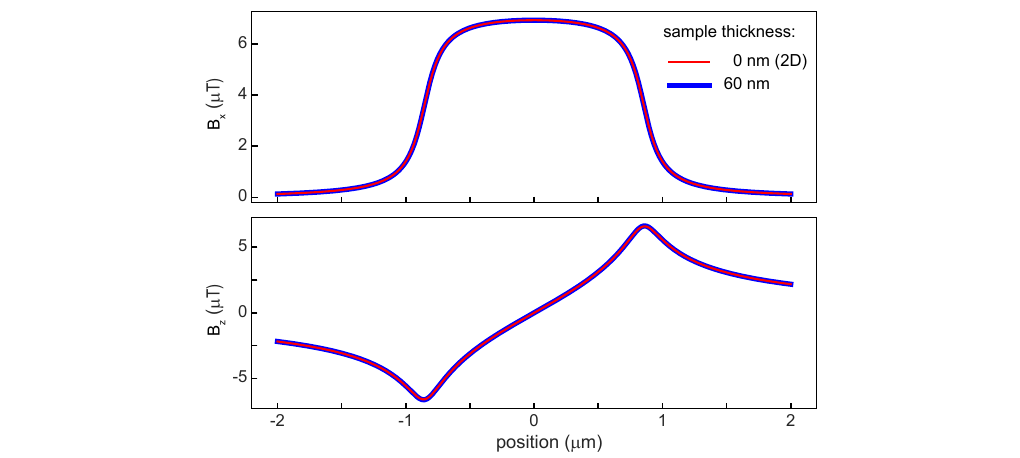}
\caption{ Spatial profiles of the $B_x$ and $B_z$ magnetic field components generated by uniform current flow at the probe height $h=140~\mathrm{nm}$ above the sample. The red line corresponds to a 2D sample with no thickness, while the blue line corresponds to a thickness of $60~\mathrm{nm}$. For the blue line, the height $h$ is measured from the center of the sample. The blue line is intentionally thicker so both curves are visible. }
\label{fig:s_thickthin}
\end{figure}

\section{Fourier magnetic field reconstruction}
The energy splitting of the NV electron spin states are sensitive to the component of the magnetic field parallel to the NV axis. 
However, under certain conditions we can extract the full magnetic field vector from such a measurement\cite{blakely_potential_1996,lima_obtaining_2009,casola_probing_2018}. 
This allowed us to reconstruct the $B_x$ component of the magnetic field, as shown in Main text Figs. 2c and 3.

The NV used in our experiment was oriented in the y-z plane. 
As the current flows along the y-axis, only the z-axis magnetic field component $B_z$ contributes field parallel to our NV. 
Thus, our NV directly measures $B_z$ as we scan above the sample, up to a factor of $1/ \sqrt{3}$ due to the angle of the NV in the y-z plane.

We measure the magnetic field in a plane above the sample. 
Assuming all sources of magnetic field are below our plane, for the magnetic field in the plane we can assume source-free magnetism: $\nabla \cdot \vec{B}=0$ and $\nabla\times\vec{B}=0$. 
These additional equations allow us to extract the full vector $\vec{B}$ from a single component. 
The full derivation is covered in the previous references, and here we focus on the special case of extracting $B_x$ from $B_z$ which was used in our measurement. The equations then reduce to the simple form:
\begin{equation}
b_x(k_x,k_y,z) = -i\frac{k_x}{\sqrt{k_x^2+k_y^2}}b_z(k_x,k_y,z)
\end{equation}
where $k_x$, $k_y$ are the $x$ and $y$ components of the planar wave vector, and $b_{x/z}(k_x,k_y,z) = \int_{-\infty}^\infty \int_{-\infty}^\infty B_{x/z}(x,y,z) \allowbreak e^{-i(k_xx+k_yy)}dxdy$ is the 2D Fourier transform of $B_{x/z}$ on the plane. 
Thus, in Fourier space $B_x$ and $B_z$ are related by a simple factor, allowing us to extract the perpendicular $B_x$ component of the magnetic field.

For our analysis, we used the profile measured along the 1D path (x-axis) above the sample, and assumed the magnetic field pattern was constant along the y-axis. Note that in Main text Fig. 2b, the magnetic field is still finite at the edge of our scanning window. 
As the Fourier integral required the field on the entire plane, this was problematic for the reconstruction. 
To improve our analysis, we extrapolated both edges of our scan with fits to $1/x$ tails before applying the reconstruction. 

\section{Electron hydrodynamics overview: Nomenclature, regimes and relevant derivations}
At high temperatures, electron transport is usually resistive.
This is in large part due to the numerous momentum-relaxing scattering events, the microscopics of which are discussed below.
Macroscopically, this leads to the well-known constitutive relation known as Ohm's `law', and can be described within the Drude theory of metals.
At very low temperatures the electron mean-free path, the average distance electrons travel in between scattering events, can exceed the geometry's length scale leading to ballistic behavior.
In certain materials, there exists a collective-transport regime in between these two limits whereby mean free paths are low, suggesting numerous scattering events, but momentum is on-average conserved following scattering.
This collective-transport behavior is commonly referred to as hydrodynamic flow due to the resemblance of the resulting flow signatures to classical hydrodynamics.
While experimentally these transport regimes are observed as a function of temperature, a more natural choice for non-dimensional independent variables is the ratio of the momentum-relaxing and momentum-conserving mean free paths to the geometry's length scale ($l_{mr}(T)/W$, $l_{mc}(T)/W$).
Note that these depend implicitly on temperature.

\subsection{Electronic Boltzmann transport equation}
In the absence of momentum-conserving scattering, the steady-state evolution of non-equilibrium electron distribution functions is given, at the relaxation time approximation (RTA) level, by:
\begin{align}
    \boldsymbol{v}_s \cdot \nabla_{\boldsymbol{r}} f(s,\boldsymbol{r}) + e \boldsymbol{E} \cdot \nabla_{s} f(s, \boldsymbol{r}) = - \frac{f(s, \boldsymbol{r})-\bar{f}(s)}{\tau_s^{mr}},
\label{eq:BTE}
\end{align}
where $f(s,\boldsymbol{r})$ is the non-equilibrium distribution of electrons around position $\boldsymbol{r}$ with state label $s$ (encapsulating the wavevector $\boldsymbol{k}$ and band index $n$). 
Non-equilibrium electrons are allowed to drift in space with a particular group velocity, $\boldsymbol{v}_s$, as a result of an external forcing term, where $e$ is the electron charge and $\boldsymbol{E}$ is an externally-applied electric field.
These drifting electrons undergo momentum-relaxing events with an lifetime, $\tau_{s}^{mr}$, which acts to returns them towards an equilibrium distribution $\bar{f}(s)$.

We investigate the flow signatures eq.~\ref{eq:BTE} permits in a two-dimensional channel, making the common assumption of a circular Fermi surface~\cite{de_jong_hydrodynamic_1995}.
Due to translational invariance along the channel, eq.~\ref{eq:BTE} simplifies to read\footnote{Note that this section uses a different convention from the main text, with the channel current flowing along the $\hat{x}$ direction}:
\begin{align}
    v_y \partial_y F(y,\theta) - e E_x v_x &= -\frac{F(y,\theta)-\langle F(y) \rangle_{\theta}}{\tau_{mr}} \\
    \sin(\theta) \partial_y F(y,\theta) + \frac{F(y,\theta)}{l_{mr}} & = e E_x \cos(\theta),
\label{eq:BTE-2}
\end{align}
where we have introduced the linearized electron deviation, $F(y,\theta)$, via $f(y,\theta) \approx f_0 - \left(\frac{\partial f_0}{\partial \epsilon} \right) F (y,\theta)$, and used $\boldsymbol{v}= v_F \binom{\cos(\theta)} {\sin(\theta)} $, $\langle F(y) \rangle_{\theta}=0$, and $l_{mr} = \tau_{mr} v_F$, where $v_F$ is the Fermi velocity, in simplifying the last line.
Finally, since eq.~\ref{eq:BTE-2} is linear, we seek normalized solutions of the form: $F(y,\theta) = e E_x \cos(\theta)\, l_{eff} (y,\theta)$ by solving:
\begin{align}
    \sin(\theta) \partial_y \, l_{eff}(y,\theta) + \frac{l_{eff}(y,\theta)}{l_{mr}} = 1,
\label{eq:BTE-3}
\end{align}
This allows a natural definition for an `average' mean free path $\tilde{l}_{eff}(y)$, which is directly proportional to current density, $j_x(y)$:
\begin{align}
    \tilde{l}_{eff}(y) &= \int_0^{2\pi} \frac{d \theta}{\pi} \cos^2(\theta) l_{eff}(y,\theta) \\
    j_x(y) &= \left(\frac{m}{\pi \hbar^2}\right) E_F e^2 \frac{E_x}{m v_F} \tilde{l}_{eff}(y)
    \label{eq:BTE-4}
\end{align}
Equation~\ref{eq:BTE-3} can be readily solved using appropriate boundary conditions, which we take to be completely diffuse, i.e.
\begin{align}z
    \forall_{\theta} \in [0,\pi) &\qquad l_{eff}(-W/2,\theta) = \frac{1}{\pi}\int_{\pi}^{2\pi}d\theta' l_{eff}(-W/2,\theta') = \mathrm{const} = 0 \\
    \forall_{\theta} \in [\pi,2\pi) &\qquad l_{eff}(W/2,\theta) = \frac{1}{\pi}\int_{0}^{\pi}d\theta' l_{eff}(W/2,\theta') = \mathrm{const} = 0,
\end{align}
where $W$ is the channel width, and the constant value at the boundary can be chosen as zero at zero magnetic field~\cite{sulpizio_visualizing_2019}.

\subsection{Ballistic to diffusive transition}
Supplementary Fig.~\ref{fig:s7}a plots the solution to eq.~\ref{eq:BTE-4} for various values of the first non-dimensional parameter, namely $l_{mr}/W$. 
It is instructive to identify the two limits.
As $\lim_{l_{mr}/W \to 0}$, i.e. the electrons undergo multiple momentum-relaxing scattering events before reaching the edge of the geometry, we recover a uniform `Ohmic' profile, with zero current density at the boundaries.
At the other extreme, as $\lim_{l_{mr}/W \to \infty}$, i.e. electrons reach the edge of the geometry without undergoing any scattering, we recover the (essentially) uniform `Knudsen' profile, with nonzero current density at the boundaries.
In between the two limits, we observe a non-diffusive current profile, which peaks when $l_{mr} \sim W$.
Note that, in contrast with the `Gurhzi' flows described below, the current density remains nonzero at the boundary.

\subsection{Hydrodynamic to diffusive transition}
Before introducing momentum-conserving terms in our electronic BTE, it is instructive to approach the problem from the opposite direction, i.e. introduce momentum-relaxing terms to a pure hydrodynamic solution.
To this end, our starting point is the electronic Stokes equation:
\begin{align}
    -\nu \partial^2_y \, j_x (y) + \tau_{mr}^{-1}\, j_x(y) = \frac{n e^2 E_x}{m},
    \label{eq:NS}
\end{align}
where $\nu= l_{mc} v_F$ is an effective electron viscosity, $n$ is the carrier density, and $m$ is the electron effective mass.
Using the (stricter) no-slip boundary conditions, i.e. $j_x(\pm W/2)=0$, the solution to eq.~\ref{eq:NS} is given by:
\begin{align}
    j_x (y) &= \frac{e^2 l_{mr}}{\hbar} \sqrt{\frac{n}{\pi}} E_x \left(1- \frac{\cosh (x/D)}{\cosh(W/2 D)}\right) \\
    &= \frac{I_{total}}{W-2 D \tanh(W/2 D)} \left(1- \frac{\cosh (x/D)}{\cosh(W/2 D)}\right),
\end{align}
where $D = \sqrt{l_{mc} l_{mr}}$ is the geometric average of the momentum-conserving and momentum-relaxing mean free paths called the `Gurzhi' length, and we have normalized all the physical constants with the total current, $I_{total}$.

\begin{figure}
\centering
\includegraphics[width=\textwidth]{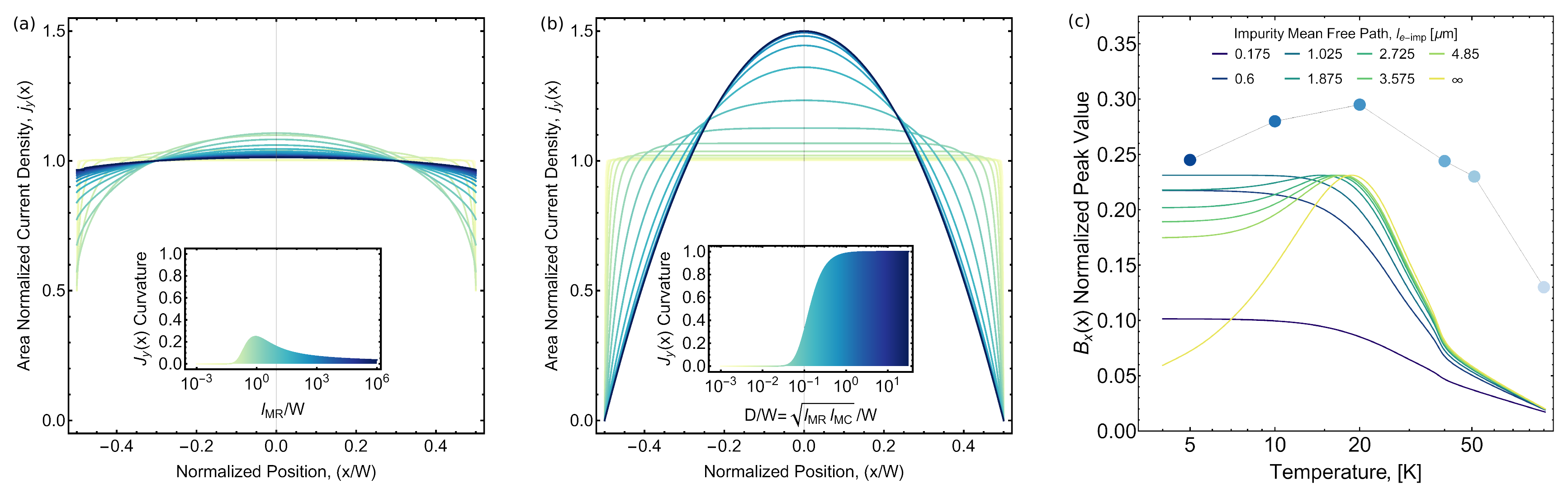}
\caption{\textbf{a} Normalized current density solutions to the momentum-relaxing electronic BTE as a function of $l_{mr}/W$, the value of which is denoted by color.
The profiles transition from uniform diffusive flow (at low $l_{mr}/W$), to non-diffusive flow (at $l_{mr} \sim W$), to ballistic `Knudsen` flow (at high $l_{mr}/W$).
This is quantified by the curvature of the current (inset), peaking at 0.25.
\textbf{b} Normalized current density solutions to the electronic Stokes as a function of $D/W$.
The profiles transition monotonically from uniform diffusive flow (at low $D/W$), to parabolic `Poiseuille' flow (at high $D/W$).
\textbf{c} Comparison of experimental profiles with the \textit{ab initio} profiles afforded within the `Knudsen` picture, showing poorer agreement as compared to Main Fig. 4c.
}
\label{fig:s7}
\end{figure}

Supplementary Fig.~\ref{fig:s7}b plots the solution to eq.~\ref{eq:NS} for various values of the non-dimensional parameter $D/W$.
As in the `Knudsen' case, as $\lim_{D/W \to 0}$ we recover a uniform `Ohmic' profile, with $j_x(y) = \frac{I_{total}}{W}$.
By contrast, as $\lim_{D/W \to \infty}$ we instead recover a fully parabolic `Poiseuille' profile, with $j_x(y) = \frac{3 I_{total}(1-4 y^2)}{2 W}$.
Note that, by virtue of our stringent boundary conditions, all solutions exhibit zero current density at the boundary and the peak current is monotonically increasing as $D/W$ increases.
Both these observations are in stark contrast with our experimental observations, and thus we turn back to the electronic BTE.

\subsection{Momentum-conserving BTE}
We seek to add additional scattering terms to eq.~\ref{eq:BTE-3}, which on-average conserve momentum:
\begin{align}
    \int_0^{2\pi}d\theta v_x \left.\frac{\partial l_{eff}(y,\theta)}{\partial t}\right|_{mc} =0
\end{align}
The simplest possible such term is given by~\cite{de_jong_hydrodynamic_1995}:
\begin{align}
    &\sin(\theta) \partial_y \, l_{eff}(y,\theta) + \frac{l_{eff}(y,\theta)}{l_{mr}} = 1 +\left(-\frac{l_{eff}}{l_{mc}} + \frac{\tilde{l}_{eff}}{l_{mc}}\right) \\
    &\sin(\theta) \partial_y \, l_{eff}(y,\theta) + \frac{l_{eff}(y,\theta)}{l} = 1 + \frac{\tilde{l}_{eff}}{l_{mc}},
\label{eq:BTE-hydro}
\end{align}
where the first additional term `depopulates' electrons according to an average mean free path $l_{mc}$ and the second term enforces these electrons are redistributed appropriately to conserve momentum.
In the last line, we use Mathhiessen's rule $l^{-1} = l_{mc}^{-1} + l_{mr}^{-1}$ to combine the mean free path terms.
Note that the presence of $\tilde{l}_{eff}$ in eq.~\ref{eq:BTE-hydro} makes our equation integro-differential.
We solve this using both an iterative and an integral solver, to avoid numerical instabilities of each method.

\subsection{Experimental comparison against ballistic limit}
Since the purely ballistic picture also permits non-diffusive flows which can exhibit non-monotonic curvature behavior, we compare the experimental results using eq.~\ref{eq:BTE-4}.
This is shown in Supplementary Fig.~\ref{fig:s7}c, which should be compared with Main Fig. 4c.
First, note that the temperature peak is shifted from $\sim15\mathrm{K}$ to $\sim20\mathrm{K}$.
Second, the non-monotonic temperature dependence requires substantially purer samples.
Finally, we note that the experimental profiles are more curved than the maximum curvature afforded by the `Knudsen` picture.

\begin{figure}
\centering
\includegraphics[width=\textwidth]{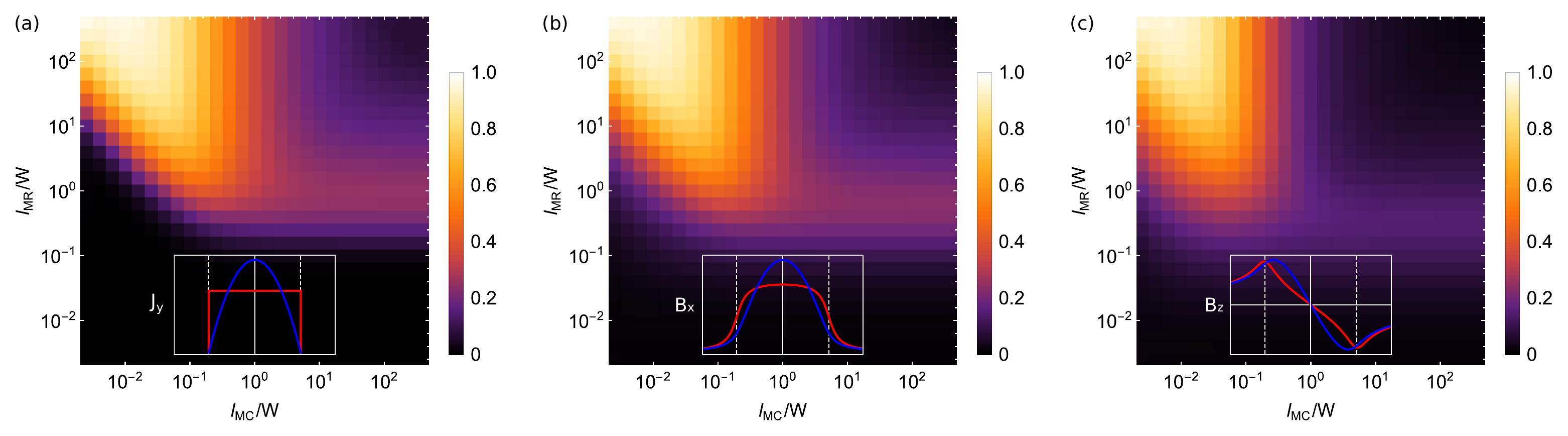}
\caption{\textbf{a} Current density curvature phase diagram for various values of $l_{mc(mr)}/W$.
Inset shows the two curvature bounds for diffusive and parabolic flows respectively.
\textbf{b} Normalized $B_x$ peak value phase diagram for various values of $l_{mc(mr)}/W$.
Inset shows the $B_x$ profiles for diffusive and parabolic flows, which acts as our normalizing lower and upper bounds respectively.
Note how the $B_x$ profile for diffusive flow is not flat, due to the finite height of the NV above the sample.
\textbf{c} Normalized $B_z$ extremum position phase diagram for various values of $l_{mc(mr)}/W$.
Inset shows the $B_z$ profiles for diffusive and parabolic flows, which acts as our normalizing lower and upper bounds respectively.
Note that for diffusive flow the $B_z$ extrema positions align with the sample edge, and move inwards for curved profiles.
}
\label{fig:s8}
\end{figure}

\subsection{Current density profiles quantification}
The solutions to eqs.~\ref{eq:BTE-4} and~\ref{eq:BTE-hydro} yield the current density as a function of position.
In Main Fig.4, we summarize our results on a two-dimensional phase diagram, quantifying the profiles using a scalar metric.
In this section, we discuss the various choices for scalar metrics and their slight differences.
Supplementary Fig.~\ref{fig:s8}a plots the current density's curvature as a function of $l_{mc(mr)}/W$.
While this is a natural choice, which is by construction normalized, we do not have direct access to the current density from our experiments.
Also note that the current density exhibits a sharp transition between the `porous' (bottom left) and hydrodynamic regimes (top left), making the differentiation of in-between regimes challenging.
By contrast, our measurements are sensitive to the magnetic field induced by this current density above the sample, and we plot the $B_x$ and $B_z$ components in Supplementary Fig.~\ref{fig:s8}b-c respectively.
The two metrics are very similar, but we prefer to use the normalized peak value of the $B_x$ profile since its connection to the current density is more intuitive.

\begin{figure}
\centering
\includegraphics[width=\textwidth]{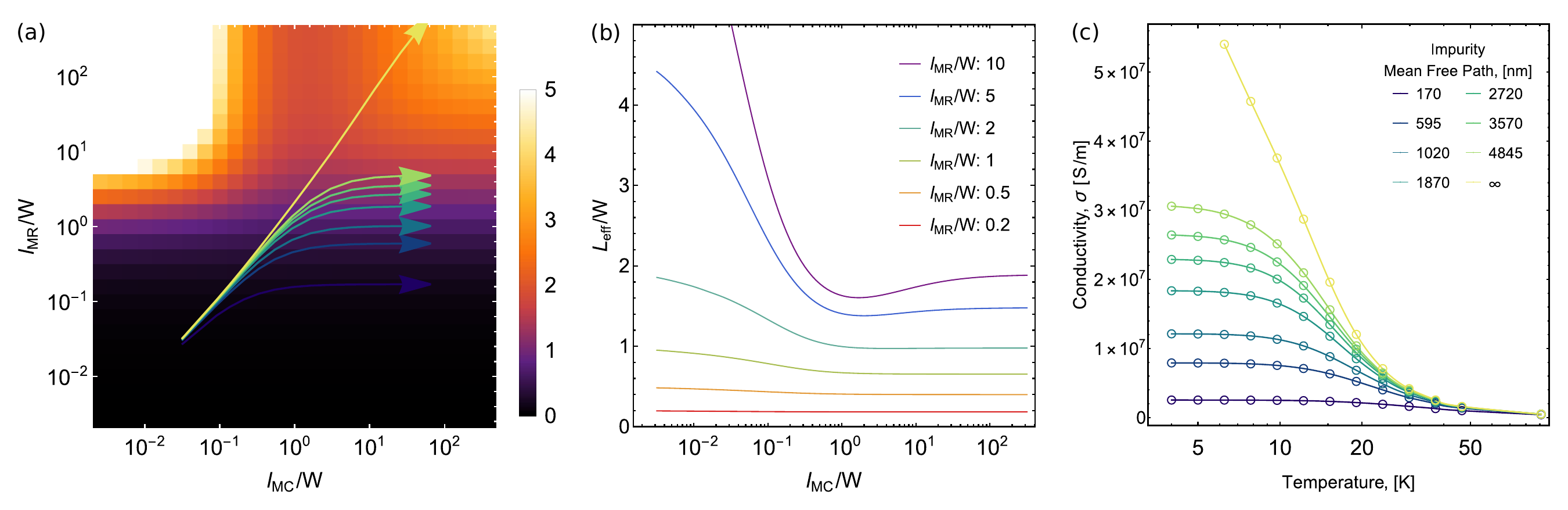}
\caption{\textbf{a} Overall effective mean free path for various values of $l_{mr(mc)}/W$.
Note the density plot is clipped at $L_{eff}/W=5$, to emphasize the regions of interest.
Overlaid arrows represent \textit{ab initio} trajectories.
\textbf{b} One-dimensional cuts of (a) for fixed values of $l_{mr}/W$, illustrating the possibility of non-monotonicity.
\textbf{c} \textit{ab initio} extracted conductivity, showing monotonic temperature dependence.}
\label{fig:s9}
\end{figure}

\subsection{Electrical conductivity}
A physical scalar representation of the current density, available using transport measurements, is the conductivity in the channel given by:
\begin{align}
    \sigma = \frac{n e^2}{m^*v_F} \int_{-W/2}^{W/2} \frac{dy}{W} \tilde{l}_{eff}(y) = \frac{n e^2}{m^*v_F}L_{eff},
\end{align}
where $L_{eff}$ is the overall effective mean free path~\cite{de_jong_hydrodynamic_1995}.
Supplementary Fig.~\ref{fig:s9}a plots $L_{eff}/W$ for various values of $l_{mr(mc)}/W$, with the \textit{ab initio} trajectories overlaid as arrows.
This indeed allows for non-monotonic behavior, highlighted by the 1D-cuts at fixed values of $l_{mr}/W$ in Supplementary Fig.~\ref{fig:s9}b.
While transport measurements of conductivity can in principle extract the non-monotonicity we observe in our spatially-resolved measurements, their dynamical range is different and in-fact our particular sample the non-monotonicity would not be observed.
This is highlighted in Supplementary Fig.~\ref{fig:s9}c, which further illustrates the importance of spatially-resolved measurements and theory.

\subsection{Anisotropic Fermi Surface}
In this section, we relax the assumption of a circular Fermi surface, used throughout the main text and preceding sections, and instead consider a `parametric` Fermi surface of the form  $\boldsymbol{k}= k_F \rho(\theta)  \binom{\cos(\theta)} {\sin(\theta)} $, where $\rho(\theta)$ is a polar equation defined between $\theta=0$ and $\theta=2\pi$, with an average value of 1. 
The Fermi velocity is then given by the vector normal to the Fermi wave-vector:
\begin{align}
    \boldsymbol{v}= v_F \binom{\cos(\theta)\rho(\theta) + \sin(\theta)\rho'(\theta)} {\sin(\theta)\rho(\theta) - \cos(\theta)\rho'(\theta)}
\end{align}

\begin{figure}
\centering
\includegraphics[width=\textwidth]{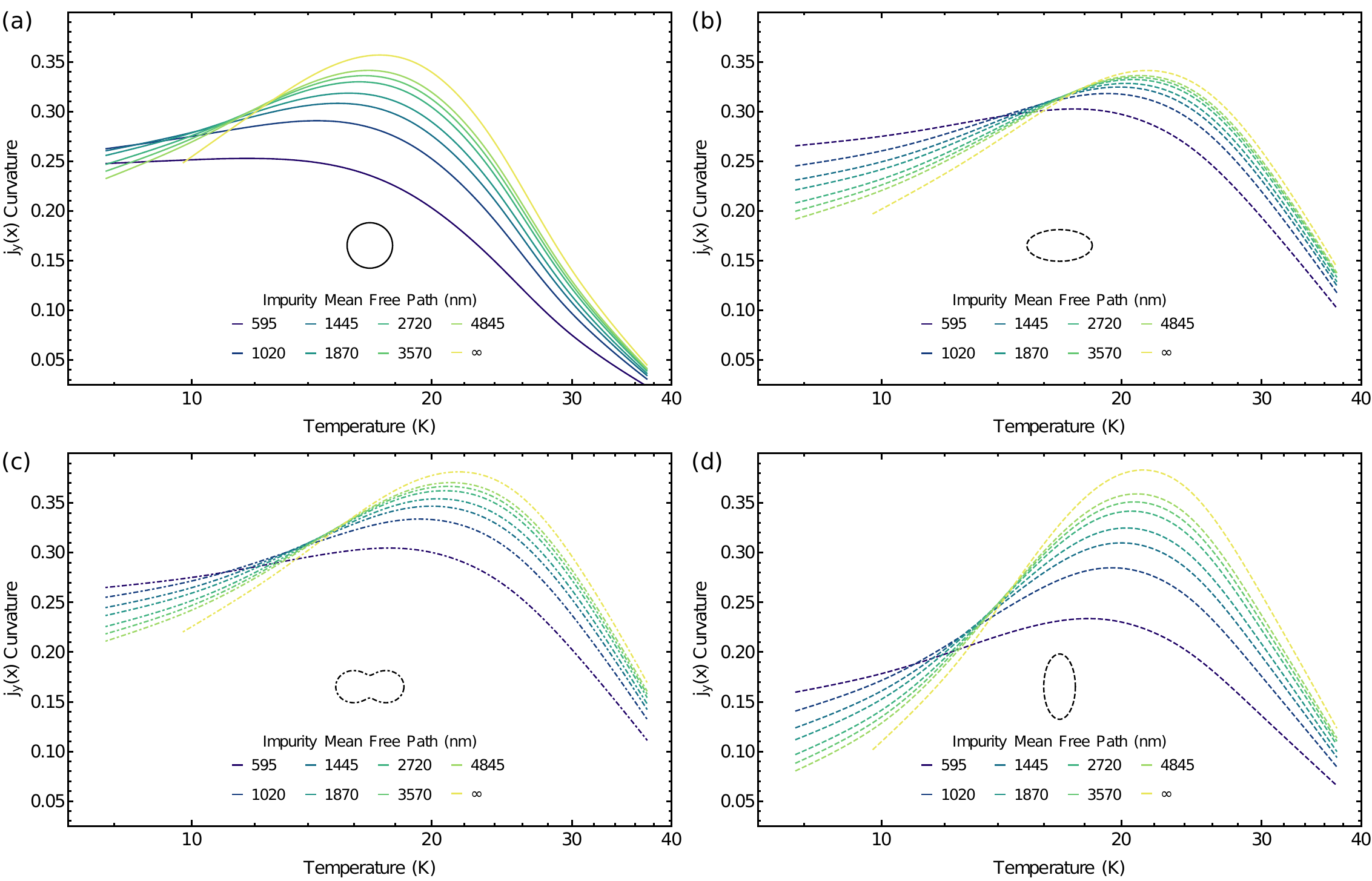}
\caption{
Current density curvature variation as a function of temperature for systems with a \textbf{a} circular, \textbf{b} elliptical (eccentricity $e=0.875$), and \textbf{c} two-fold anisotropic (anisotropy strength $\epsilon=0.5$) Fermi surfaces.
\textbf{d} Comparison between the relative of orientation of the Fermi surface to the channel direction for an elliptical Fermi surface with the same eccentricity as in \textbf{b}.}
\label{fig:sX}
\end{figure}

Note that we naturally recover the circular Fermi surface limit by setting $\rho(\theta)=1$.
In particular we consider elliptical Fermi surfaces with an eccentricity $e$ and two-fold parametric Fermi surfaces with an anisotropy strength $\epsilon$, rotated at an angle $\theta_0$ with the horizontal axis:
\begin{align}
    \rho_{\mathrm{ellipse}}(\theta)&=\sqrt{2}\sqrt{-\frac{\sqrt{1-e^2}}{-2+e^2+e^2 \cos[2(\theta-\theta_0)]}}\\
    \rho_{\mathrm{two-fold}}(\theta)&=1+\epsilon \cos[2(\theta-\theta_0)]
\end{align}
The derivation of the anisotropic BTE proceeds by generalizing the parametrization in the main text $F(y,\theta) = e E_x v_x(\theta)\, l_{eff} (y,\theta)$ and using the fact that the drift-velocity along the transverse direction is zero at steady-state to decouple the terms in the integral~\cite{de_jong_hydrodynamic_1995}. 
This leads to an equation of the form:
\begin{align}
    \hat{v}_y(\theta) \partial_y \, l_{eff}(y,\theta) + \frac{l_{eff}(y,\theta)}{l} = 1 + \frac{1}{l_{mc}}\int_0^{2\pi} \frac{d \theta'}{\pi} \hat{v}_x^2(\theta) l_{eff}(y,\theta'),
\label{eq:BTE-anisotopic}
\end{align}
which we solve numerically using an iterative finite-element solver. 
Equation~\ref{eq:BTE-anisotopic} is convention-dominated, which presents numerical difficulties for certain parametric Fermi surfaces exhibiting $\theta$-discontinuities.
We address this by adding a small amount ($\approx 5\times 10^{-4}$) of artificial diffusion to the right-hand-side of eq.~\ref{eq:BTE-anisotopic}.

Figure~\ref{fig:sX} plots the current density curvature as a function of temperature for various empirical impurity values (color-code), for different Fermi surface parametrizations.
While the temperature non-monotonicity is perserved in all these configurations, with a peak around $15-25$K, the exact position of the peak appears to shift to higher temperatures for non-spherical Fermi surfaces.
The effects of introducing both significant anisotropy in the Fermi surface (Fig.~\ref{fig:sX}b) as well as non-convex regions (Fig.~\ref{fig:sX}c) on the range of curvature values achieved appear to be otherwise weak.
Finally, we investigated the effect of Fermi surface alignment with the channel direction (Fig.~\ref{fig:sX}d).
The increased distribution of Fermi velocity directions aligned with the channel can indeed lead to increased values of curvature~\cite{bachmann2021directional}.
While this phenomenon is most pronnounced in the ballistic limit~\cite{bachmann2021directional}, we see remnants of this behavior persist in the regime we're operating in - with considerable momentum-conserving interactions.

\section{Computational details}
The \textit{ab initio} calculations were performed with the open source density functional theory (DFT) code JDFTx~\cite{sundararaman2017jdftx}.
First we fully relaxed the $T_d$-WTe$_2$ (Fig.~1c, main text) using fully relativistic Perdew-Burke-Ernzerhof pseudopotentials~\cite{perdew1996generalized,DALCORSO2014337,Perdew2008restoring} and Grimme's D-2 van der Waals approach~\cite{grimme2006semiempirical}. 
A kinetic cutoff energy of 40~Ha was used along with a $14\times7\times4$ Gamma-centered $k$-mesh, and a Fermi-Dirac smearing of 0.01~Ha for the Brillouin zone integration. 
Both the lattice constants and the ion positions were relaxed until the forces on all atoms were less than 10$^{-8}$~Ha/Bohr. 
The relaxed lattice constants were found to be $a = 3.46$~\AA, $b = 6.20$~\AA, and $c=13.09$~\AA, respectively. 
To compute the e-ph scattering time, we performed frozen phonon calculations in a $1\times1\times1$ supercell, and obtained 88 maximally localized Wannier functions (MLWFs) by projecting the plane-wave bandstructure to W $d$ and Te $p$ orbitals, which allowed us to converge the electron scattering calculation on a much finer $112\times56\times32$ ($224\times112\times64$) $\mathbf{k'}$ and $\mathbf{q}$ grid for $T>(<)20$~K. 
{\taueeph} was collected on 56 irreducible \textbf{q} points in the Brillouin zone. 
For e-e scattering, we reduced the $k$-mesh to $6\times3\times2$ due to the computational cost, for which we verified the electronic structure.  
A dielectric matrix cutoff of 120~eV was used to include enough empty states, with a energy resolution of 0.01~eV.

Further, since the electronic structure of $T_d$-WTe$_2$ is known to be sensitive to lattice strain~\cite{soluyanov_type-ii_2015}, the Weyl semimetal phase (WSM) was realized by confining the lattice constants to $a = 3.48$~\AA, $b = 6.25$~\AA, and $c=14.02$~\AA. 
Here we employed fully relativistic Perdew-Wang~\cite{perdew-wang} pseudopotentials with the same settings as described earlier to obtain stable phonons. 
An extensive comparison between the \emph{Relaxed} phase and \emph{WSM} phase is shown in Sec.~10.

\section{Relaxed phase vs. Weyl semimetal phase}
\begin{figure}
    \centering
    \includegraphics[width=0.9\textwidth]{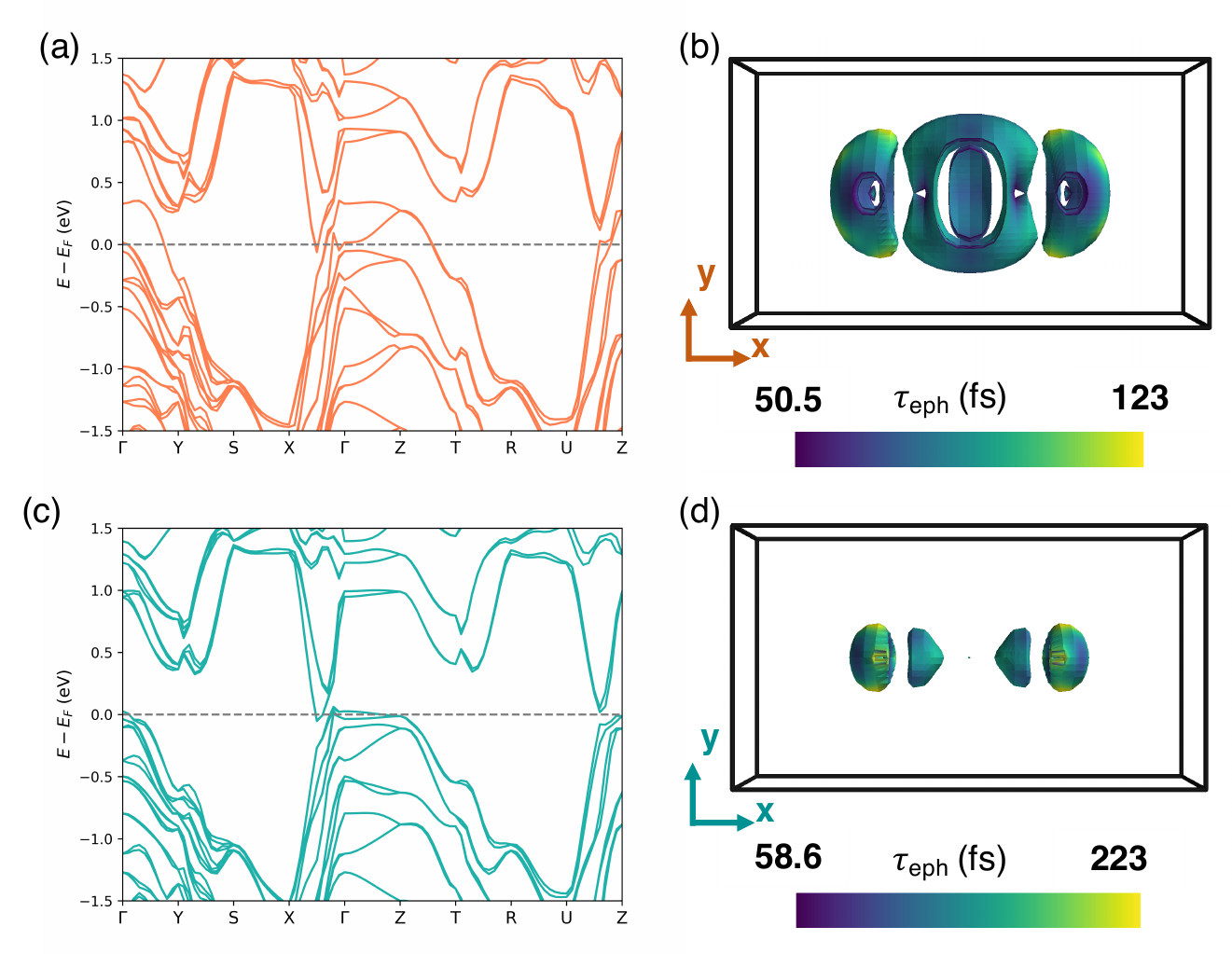}
    \caption{Electronic bandstructures and Fermi surfaces for relaxed (a-b) and WSM (c-d) WTe$_2$ phases. In (b) and (d) we show the  electron-phonon lifetimes at 28~K projected onto the Fermi surfaces to highlight the anisotropic feature.}
    \label{fig:s10}
\end{figure}

As mentioned in the previous section, the WSM phase has $\sim7$~\% tensile strain along the stacking $\hat{z}$ axis. 
The phonon modes only show slight softening along the $\Gamma-Z$ direction compared to the relaxed structure, without significant difference overall. 
However, the relaxed phase has much more dispersive electronic structure along $\Gamma-Z$ direction due to a shorter distance between the van der Waals layers (Supplementary Fig.~\ref{fig:s10}(a) and (c)).
Consequently, the Fermi energy cuts through these bands, leading to a slightly more `metallic' behavior in the relaxed phase. 
One may expect leaving a larger hole pocket (Supplementary Fig.~\ref{fig:s10}(b) and (d)).
However, when we examine the spatially resolved momentum relaxing electron-phonon lifetimes on the Fermi surface in these two structures, the WSM phase features longer-lived electrons than the relaxed phase.

To better understand their hydrodynamic behavior, we performed the same numerical calculation with \emph{ab initio} $\tau_{\rm ee}$, $\tau_{\rm ee}^{\rm PH}$, and $\tau_{\rm eph}$. 
We found that while $\tau_{\rm mr}$ does not show predominant distinction, $\tau_{\rm ee}$ is greatly decreased in the WSM phase. 
As a result, the non-monotonic temperature dependence of the resulting current density profiles peaks at $\sim8\mathrm{K}$, as shown in Supplementary Fig.~\ref{fig:s11}.

\begin{figure}
    \centering
    \includegraphics[width=0.75\textwidth]{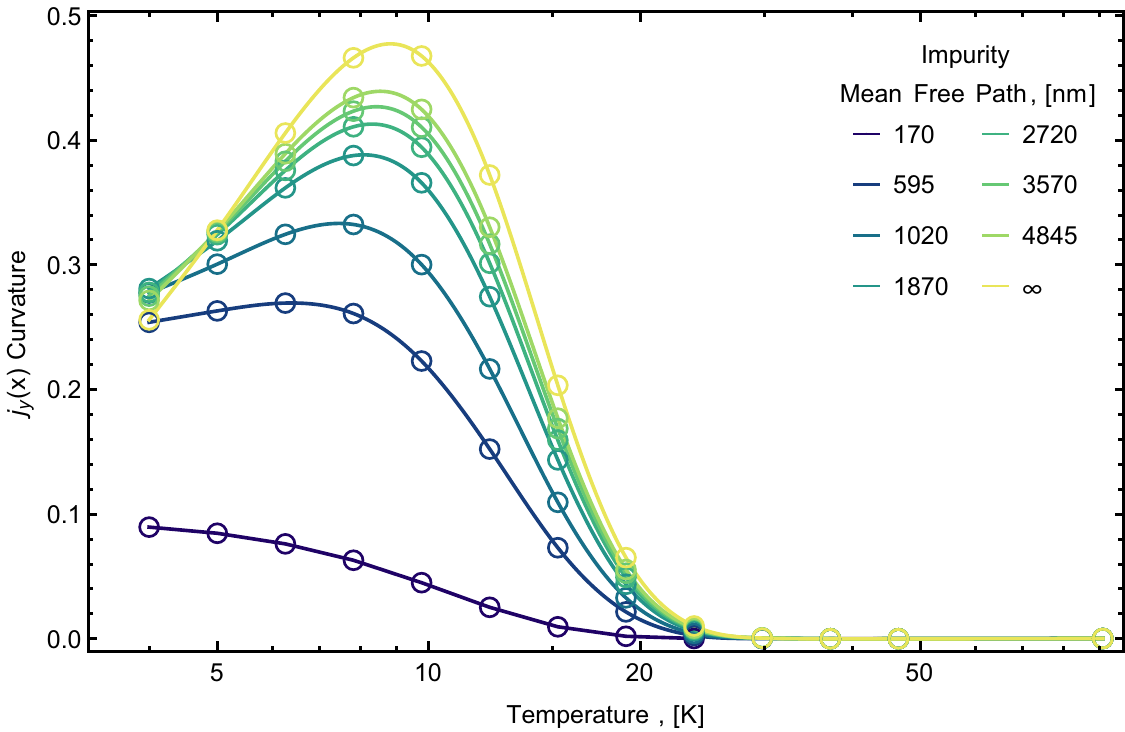}
    \caption{\textit{Ab initio} current density curvature solutions as a function of temperature for the Weyl semimetal phase.
    Note, the peak occurs at a lower temperature, and the curvature values are significantly higher than those observed experimentally.
    }
    \label{fig:s11}
\end{figure}

\section{Discussion on high temperature deviations between predicted behavior and experimental observations}
As shown in Main Figs.~3,4, the experimental observations above $\sim40\mathrm{K}$ deviate significantly from the predicted behavior.
While the theoretical predictions admit almost entirely diffusive profiles, experimentally we observe curved profiles.
In this section, we expand on possible causes for this discrepancy, which can be attributed to approximations made by the theory and experimental uncertainties.
First, oxidation effects in the WTe$_2$ flake result in a non-uniform distribution of impurities, which in turn suggests a position-dependent mean free path.
It is instructive to consider the extreme case in which such impurities permit no current density at the edges, effectively reducing the sample width, artificially suggesting enhanced flow in the center.
Allowing for $\sim10\%$ change in the effective width, for example, accounts for the discrepancy at $90~\mathrm{K}$.
Similarly, while the theoretical model assumes an infinite two-dimensional ribbon, the sample has both finite length ($\sim100~\mathrm{\mu m}$) and finite thickness ($\sim50~\mathrm{nm}$).
Since the electron mean free paths (Main Fig.~4a) are of the same order as the sample thickness, this will result in a transition between `Knudsen' and `Poiseuille' flow along the $\hat{z}$ direction, modifying the profiles along the $\hat{x}$ direction.
Finally, since our sample is a thin WTe$_2$ flake exfoliated on a quartz substrate, the phonon spectrum will likely be modified away from that of the bulk crystal.
Furthermore, considering the entire WTe$_2$ phonon spectrum ($< 250 \mathrm{cm}^{-1}$) is contained within the quartz phonon spectrum, bulk phonons from the WTe$_2$ can scatter into the substrate with little interface resistance, thus providing another competing scattering mechanism. 

\section{Electron-Hole Scattering Channels}
\begin{figure}[!ht]
    \centering
    \includegraphics[width=0.5\textwidth]{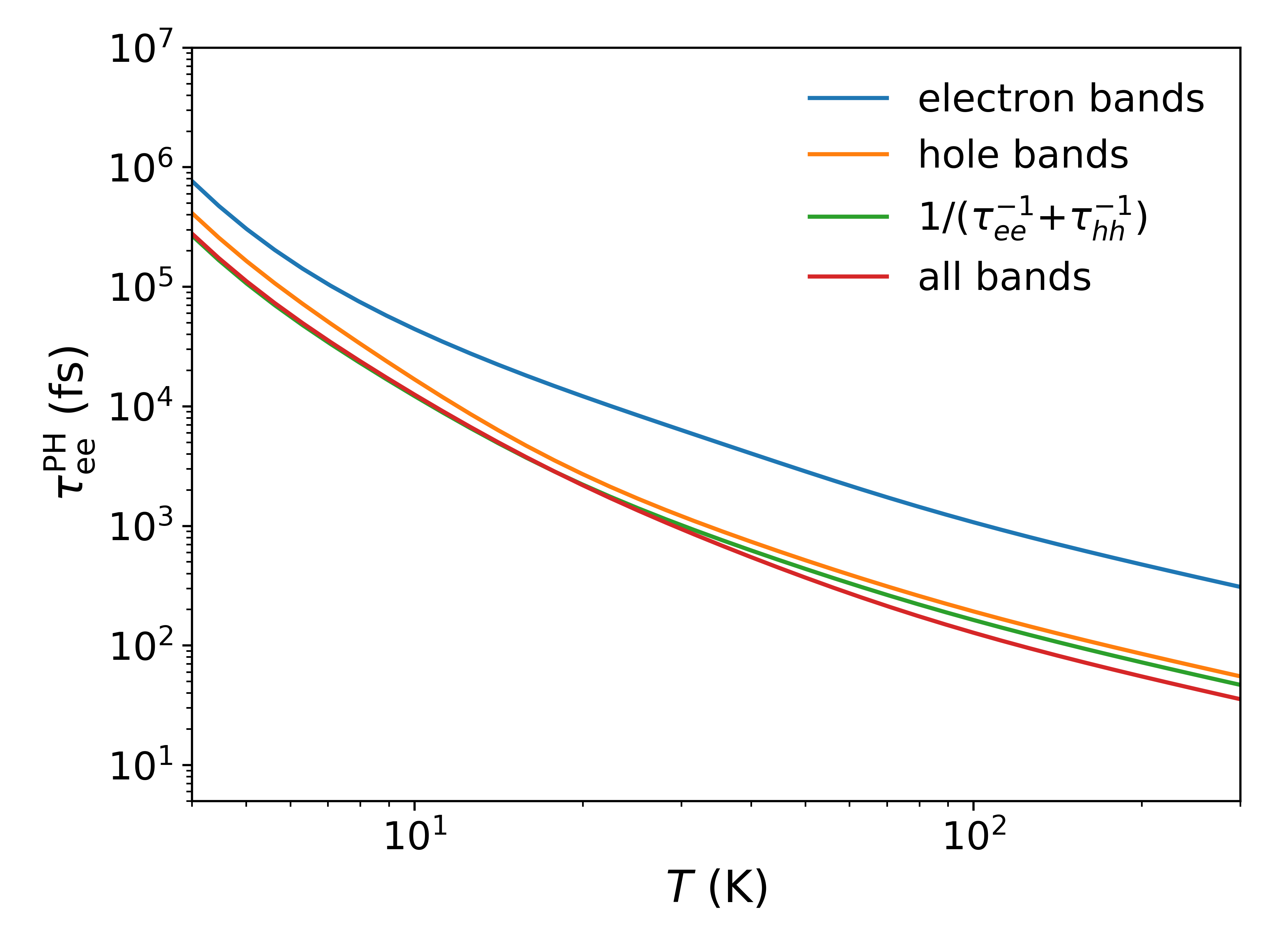}
    \caption{Phonon mediated carrier lifetimes at different temperatures, computed with all electron bands (blue), all hole bands (orange), and all bands (red), while the green curve is calculated by summing the electron and hole contributions using Matthiessen's rule.}
    \label{fig:tau_ee_hh}
\end{figure}

As indicated by magnetoresistance and Hall resitivity signatures (Figs.~S1-S3), WTe$_2$ is a semimetal with both electron and hole carriers present. 
In this section we discuss the effect of scattering between the electron and hole pockets on the phonon-mediated time scale ($\tau_{\rm ee}^{\rm PH}$).
The identification of different charge carriers in the \textit{ab initio} calculations is enabled by the well-defined band index used in the maximally localised Wannier functions (MLWFs) formalism, and subsequently adopted in our calculation of the electron phonon matrix elements.
To this end, we compute the phonon mediated carrier lifetimes allowing scattering between only the electron (blue) or hole (orange) pockets (Fig.~\ref{fig:tau_ee_hh}), and compare these against the lifetime values obtained allowing for scattering between all bands (red).
We note that both momentum and current is conserved for electron-electron and hole-hole scattering, while only momentum is conserved for electron-hole scattering.
To address the extent of electron-hole scattering, we use Matthiessen's rule to sum the contributions assuming electron-electron and hole-hole scattering as independent channels (green curve in Fig.~\ref{fig:tau_ee_hh}).
The deviation of the effective lifetime given by Matthiessen's rule (green) and the computed lifetime using all bands (red) is an indication of how independent these scattering channels are.
At the temperatures of interest in this work ($\mathrm{T} \lesssim 90\mathrm{K}$), the two scattering channels are almost entirely independent, and phonon-mediated scattering between electron and hole bands can be neglected.

\section{Electron-phonon coupling strength in {WT\lowercase{e}$_2$}}
In this section we calculate the dimensionless coupling strength in WTe$_2$, which is closely related to the estimation of the critical temperature in phonon-mediated superconductors following the BCS theory~\cite{PhysRev.167.331}, to facilitate further comparison of WTe$_2$ with other systems.
The electron-phonon coupling strength associated with a specific phonon mode (\textbf{q}$\nu$) can be calculated by 
\begin{equation}
    \lambda_{\textbf{q}\nu}=\frac{1}{N(\varepsilon_F)\omega_{\textbf{q}\nu}}\sum_{mn}\int\frac{\mathrm{d}\textbf{k}}{\Omega_{\rm BZ}}|g_{mn,\nu}(\textbf{k,q})|^2 \times \delta(\varepsilon_{n\textbf{k}}-\varepsilon_F)\delta(\varepsilon_{m\textbf{k+q}}-\varepsilon_F),
\end{equation}
where $N(\varepsilon_F)$ is the density of states per spin at the Fermi level and other symbols follow the same definition as in the \emph{Methods} session. 
The isotropic dimensionless electron-phonon coupling strength ($\lambda$) of WTe$_2$ is given by the Brillouin zone sum over a uniform $14\times 7\times 4$ \textbf{q}-grid to be \textbf{0.49}.
As means of comparison with a known BCS superconducting system, we obtained $\lambda = \mathbf{0.745}$ on a $12\times 12\times 12$ \textbf{q}-grid for aluminum (superconducting at $T_C\approx1.75$~K), while $\lambda = \mathbf{0.17}$ on a $8\times 8\times 8$ \textbf{q}-grid for copper (non-superconducting).

\section{Carrier density effect on the scattering length scales in {WT\lowercase{e}$_2$}}
\begin{figure}[!ht]
        \centering
        \includegraphics[width=\textwidth]{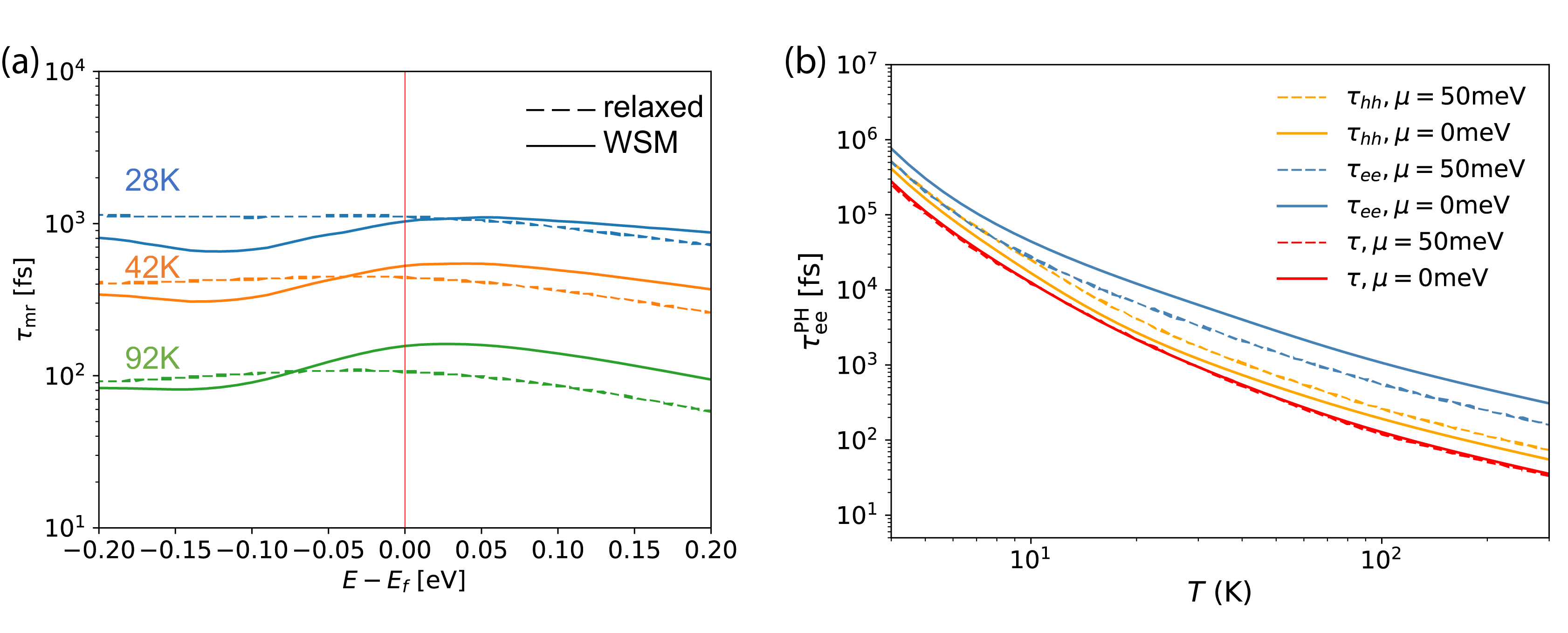}
        \caption{(a) Momentum relaxing lifetime ($\tau_{\rm mr}$) at different electron chemical potentials at various temperatures (color coded), in both the relaxed structure (dashed lines) and the WSM phase (solid lines). 
        (b) Phonon-mediated carrier lifetimes at different temperatures with a chemical potential shift of 50~meV, computed with all electron bands (blue), all hole bands (orange), and all bands (red).}
        \label{fig:taumr_mu}
    \end{figure}

In layered ver der Waals structures, the Fermi level and thus the carrier density, are sensitive to structural characteristics such as crystal stoichiometry, strain, defects etc. 
In this section we discuss how differences in carrier density affect the momentum relaxing mean free paths ($l_{mr}$) in WTe$_2$ through \emph{ab initio} calculations of both momentum relaxing and momentum conversing length scales through changing the electron chemical potential.
For the momentum relaxing length scale, we explicitly examine, how the $\tau_{\rm mr}$ changes at different electron chemical potential, or Fermi level, in WTe$_2$ for both the WSM phase (solid lines) and the relaxed structure (dashed lines) at multiple temperatures shown in Fig.~\ref{fig:taumr_mu}a.
Allowing for a chemical potential range within $\pm 200$~meV, we do not see a significant change of $\tau_{\rm mr}$.
For the momentum conserving length scale, in this context the phonon-mediated carrier mean free path, the carrier density could affect the phase space for phonon-electron interaction as well as the Brillouin zone integration of ($\tau_{n\textbf{k}}$) since only the states near the Fermi level are required, typically weighted by $-f'(\varepsilon_{n\textbf{k}})$ under the Boltzmann relaxation time approximation (RTA). 
As shown in Fig.~\ref{fig:taumr_mu}b, while the contribution from the electron (blue) and hole (orange) carriers could be modified when shifting the chemical potential above the Fermi energy by 50~meV, the change in the overall phonon-mediated length scales (red) is negligible. 
Therefore, we believe the carrier density will not hinder the intrinsic hydrodynamic limit in WTe$_2$.

\bibliography{wte2_hydro_citations}